\documentclass{article}
\usepackage{graphicx} 
\usepackage{amsmath} 
\usepackage{amssymb}
\usepackage{bm}
\usepackage{float}
\usepackage{comment}
\usepackage{cite}
\usepackage{xcolor}
\newcommand{\rev}[1]{{\color{black}#1}}
\usepackage[english]{babel} 

\title{Interaction Landscapes of Structurally Distorted Toroidal Classical Spin Textures}
\author{
Henry Pommer$^{1,\dagger}$,
Arkamita Bandyopadhyay$^{1,\dagger,*}$,
Jamal Berakdar$^{1,*}$
}

\begin{document}

\selectlanguage{english} 

\maketitle
\begin{center}
\small
$^{1}$Institut f\"ur Physik, Martin-Luther Universit\"at Halle-Wittenberg,
and  Halle-Berlin-Regensburg Cluster of Excellence,
Center for Chiral Electronics,
06099 Halle/Saale, Germany\\
$^{\dagger}$These authors contributed equally to this work.\\
$^{*}$Corresponding authors:
arkamita.bandyopadhyay@physik.uni-halle.de,
jamal.berakder@physik.uni-halle.de
\end{center}
\begin{abstract}
We develop a classical discrete magnetostatic model of localized magnetic moments arranged in a toroidal texture, where the moments are represented by static current loops and study  the behavior of  the magnetic dipole-dipole interaction under several types of structural deformations.  The purpose is 
to disentangle effects due to  geometric and magnetic deformations, including elliptical, lopsided, alternating, and tilted configurations. In particular, we investigate how  distortions modify the toroidal moment and the interaction energy between pairs of toroidal structures. The resulting interaction landscapes exhibit characteristic symmetry patterns and anisotropies that are highly sensitive to the underlying deformation. As a separate proof-of-concept comparison, first-principles calculations on a molecular spin-toroidal system reveal qualitatively similar orientation-dependent behavior. \rev{The calculated interaction is microscopically the conventional pairwise magnetic dipole-dipole interaction between localized moments arranged in a spin-toroidal texture; the toroidal moment characterizes the collective texture and should not be interpreted as introducing an additional elementary interaction term.} Our results provide a  platform for exploring \rev{orientation-dependent interactions of spin-toroidal textures} in molecular magnets and related nanoscale magnetic systems that are essential to understand the role of toroidal order.
\end{abstract}
\section{Introduction}
A classical charge $q_i$  and current density  $\mathbf{J}_i$  distribution may have, in addition to an electric $\boldsymbol{\mu}_{j}$ and magnetic dipoles moments $ \mathbf{m}_{i}$, also a toroidal  dipole moment $\mathbf{T}$. 
For example, $\mathbf{T}$ is finite for  vortex-like arrangements of magnetic moments $ \mathbf{m}_{i}$ 
\cite{dubovik1990toroid,kopaev2009toroidal,spaldin2008toroidal} and transforms  under space and time inversion in a different way than  $ \mathbf{m}_{i}$  and  $ \boldsymbol{\mu}_{j}$ , as clear from its definition 
\begin{equation}
\mathbf{T}
=
\frac{1}{2}
\sum_{i}
\mathbf{r}_{i}
\times
\mathbf{m}_{i}=\frac{1}{2}
\sum_{ij}
q_j\mathbf{r}_{j}
\times
\mathbf{m}_{i}\delta_{ij}=\frac{1}{2}
\sum_{ij}
\boldsymbol{\mu}_{j}
\times
\mathbf{m}_{i}\ \delta_{ij}
\end{equation}
where $\mathbf{r}_{i}$ and $\mathbf{m}_{i}=\frac{1}{2} \sum_i \mathbf{r}_i\times \mathbf{J}_i $.  For insulating spin-based systems and neglecting quantum fluctuations, the magnetic moments are proportional to the underlying local spin vectors, $\mathbf{m}_i \propto \mathbf{s}_i$.  An emergent  toroidal moment  is directly related to the spin texture  in this case. An important caveat is that the total toroidal moment can be sizable  even when the net magnetic moment vanishes.Thus, toroidal order provides a distinct feature of magnetic textures beyond conventional collinear magnetism and has attracted considerable interest in areas ranging from magnetoelectric materials and metamaterials to molecular magnetism and spin-based information processing.\cite{spaldin2008toroidal,ahmadivand2020toroidal, pister2022toroidal,zhang2020single}
The macroscopic classical origin of the toroidal moment and the toroidal orders have been studied extensively since the early 2000s, with the open question about the microscopic origins of ferrotoroidic ordering \cite{spaldin2008toroidal, ederer2009toroidal, rabe2007response}. In recent years, studies have shown that, at the microscopic level, toroidal moments naturally arise from closed arrangements of localized magnetic moments. Such spin-toroidal textures have been identified in a variety of systems, including lanthanide-based molecular clusters,\cite{ungur2014single,baniodeh2015ligand, yang2022toroidal, das2015single, guo2012first,sharma2025design}, ferrotoroidal complexes,\cite{chauhan2025ultraslow, del2019organometallic} self-assembled toroidal nanostructures,\cite{kim2013development} and molecular magnets\cite{li2019recent, wang2022ferroelectric}. In these systems, the magnetic moments form vortex-like configurations that support a finite toroidal moment while often maintaining a nearly vanishing net magnetic dipole moment. These lanthanide-based molecular complexes serve as a leading platform for realizing toroidal states owing to their strong magnetic anisotropy arising from large orbital angular momentum and spin–orbit coupling.\cite{chauhan2025ultraslow} The stabilization of toroidal order relies critically on the orientation of local anisotropy axes, motivating investigations into how geometric distortions and deviations from ideal symmetry influence toroidal behavior. The discovery of field-controlled ferritoroidal materials and ferrotoroidal phases in chiral magnets has renewed interest in understanding the microscopic mechanisms governing toroidal order and its stability~\cite{ding2021field}. Toroidal geometries similar to the idealized structure considered here have previously been investigated in molecular nanotorus systems, where curvature and topology give rise to unique electronic and magnetic properties \cite{sarapat2019review, carbonNanotori}. Recent works have further highlighted the importance of spin-toroidal systems by demonstrating novel toroidal spin textures, such as homochiral toroidal spin states in Dy(III)-based single-molecule toroics, providing a promising platform for magnetoelectric coupling, chiral spintronics, and molecular information technologies~\cite{zhu2026homochiral, pister2022toroidal,zhang2020single,hymas2025preparation}.\\
While substantial progress has been made in understanding the origin, symmetry properties, and experimental signatures of toroidal moments in both bulk ferrotoroidic materials and single-molecular tori, considerably less attention has been devoted to the interaction between individual toroidal units. Existing studies primarily focus on the formation and detection of toroidal ground states, whereas the influence of geometry on \rev{the dipole-dipole interaction between spin-toroidal textures} remains largely unexplored. In particular, it is not yet clear how structural distortions modify the interaction landscape between neighboring toroidal structures or whether the characteristic \rev{symmetry signatures of the underlying toroidal texture} survive away from ideal geometries. \\
Assuming  that in a magnetic textures, individual building blocks  have established equilibrium toroidal state, and if  spin fluctuations are marginal, (e.g.\ when the localized magnetic moments are large) then the question arises which global states are energetically  more favorable. 
Our  first step to approach this issue  is  a classical model consisting of discrete magnetostatic  elements. As detailed below the model allow
for studying how geometry and magnetic-moment orientation affect toroidal interaction landscapes. The localized magnetic moments are treated as classical magnetic dipoles. For for orbitally induced magnetism each magnetic moment can be viewed to be at the center of a static current loop.  The mutual magnetic interactions are evaluated through the magnetic dipole–dipole interaction. While no quantum spin Hamiltonian or quantum dynamics is explicitly included, the model captures the geometric and magnetostatic consequences of vortex-like arrangements of localized magnetic moments and therefore provides a physically transparent description of interaction trends relevant to spin-toroidal molecular and nanoscale systems.
A classical description is particularly useful here because it provides a physically transparent framework in which geometry, symmetry, and the spatial arrangement of local magnetic moments can be varied independently. This allows the resulting changes in the toroidal moment and interaction landscape to be associated directly with structural or magnetic-texture modifications, without simultaneously introducing material-specific effects such as exchange interactions, ligand-field contributions, or detailed electronic structure. The model therefore provides a simple reference framework for identifying symmetry relations, angular dependencies, and qualitative interaction trends that can also be relevant to more complex molecular and nanoscale spin-toroidal systems.\\
Here we investigate the consequences of elliptical and lopsided geometric deformations, as well as tilted spin configurations. We  present analytical expressions for the toroidal moment, and based on that the resulting interaction landscapes are analyzed within a pairwise dipole-dipole interaction model. We also discuss  the orientations in which symmetry breaking modifies the angular dependence of the interaction energies of a pair of tori. We show that geometric distortions or spin tilting generate pronounced changes in the interaction landscape, producing new minima, maxima, and anisotropic features that are absent in ideal toroidal geometry. \\
Finally, we compare the predictions of the discrete model with an experimentally realized Tb$_4$ molecular toroidal system\cite{yang2022toroidal} as a proof-of-concept example. Although realistic molecular systems involve additional electronic and structural contributions beyond pure toroidal coupling, the observed orientation-dependent interaction energies exhibit trends that are qualitatively consistent with those predicted by the discrete spin-toroidal framework. The present study therefore provides a unified description of how geometry and spin configuration govern toroidal interactions and offers insight into the design of future molecular and nanoscale toroidal systems.
\section{Discrete Spin-Toroidal Model}
\subsection{Ideal Toroidal Geometry}
\begin{figure}[H]
    \centering
    \includegraphics[width=0.95\linewidth]{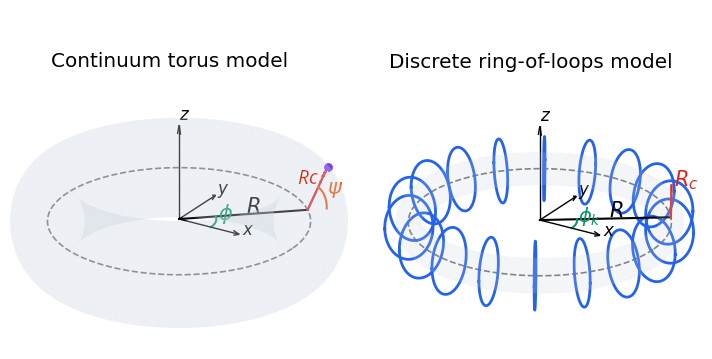}
    \caption{
     Continuum and discrete descriptions of an ideal toroidal structure. The continuum torus is characterized by major radius $R$ and minor radius $R_c$, while the discrete model consists of localized magnetic moments arranged on current loops distributed around the toroidal ring. The resulting vortex-like magnetic texture supports a finite toroidal moment $\mathbf{T}$ and serves as the reference geometry for the deformation studies presented in this work.
    }
    \label{fig:torus}
\end{figure}

Our goal is to develop a simple classical model of a toroidal arrangement of localized magnetic moments and to study how its toroidal dipole moment responds to geometric and physical deformations. The discrete formulation naturally allows for distortions such as nonuniform loop sizes, imperfect ordering, and lopsided toroidal geometries. The localized magnetic moments considered here may be viewed as arising from small current loops distributed along the toroidal ring, providing a microscopic realization of the continuum torus shown in Fig. \ref{fig:torus}.
This controlled construction makes it possible to identify how individual geometric or magnetic-texture modifications affect the toroidal response before introducing the additional microscopic interactions present in a specific material.
We consider a toroidal structure composed of \(N\) localized magnetic moments arranged uniformly on a ring of major radius \(R\), with each moment associated with a local loop of radius \(R_c\) (Fig.~\ref{fig:torus}). The discrete-loop representation serves as a geometrical description of a vortex-like spin texture and provides a natural framework for introducing geometric distortions.
For the ideal toroidal configuration, the toroidal moment is
while the net magnetic dipole moment vanishes. Here \(I\) denotes the effective current associated with each local loop and the magnetic moment on each loop  becomes
\begin{equation}
\mathbf{m}(\phi)
=
I\pi R_c^2\,\hat{\mathbf t}(\phi).
\end{equation} with \(\hat{\mathbf t}(\phi)\) denoting the local tangential direction along the toroidal ring. This configuration serves as the reference state for the deformation studies presented below.
\subsection{Deformation I: elliptical ring of loop centers}
We now deform the major ring from a circle to an ellipse, where the center of the $k$-th loop is taken as
\begin{equation}
\mathbf{r}_k
=
(R_x\cos\phi_k,\,
R_y\sin\phi_k,\,
0).
\label{eq:ellipsecenters}
\end{equation}
and the toroidal moment becomes (see Supporting Information for details),
\begin{equation}
\mathbf{T}
=
\frac{1}{2}
\sum_{k=0}^{N-1}
\frac{I\pi R_c^2 R_xR_y}
{\sqrt{R_x^2\sin^2\phi_k+R_y^2\cos^2\phi_k}}
\,\hat{\mathbf z}.
\label{eq:Tellipse}
\end{equation}
Where, the toroidal moment remains along $\hat{\mathbf z}$, but its magnitude is modified by the breaking of rotational symmetry.
\subsection{Deformation II: position-dependent poloidal radius}
We next consider a toroidal structure in which the radii of the local loops vary along the ring (see Fig. ~\ref{fig:disttor}). Such a deformation provides a simple description of a lopsided torus with non-uniform loop sizes while preserving the overall toroidal topology. The magnetic moment associated with the \(k\)-th loop is taken as
\begin{equation}
\mathbf{m}_k = I\pi R_c^2(\phi_k)\,\hat{\boldsymbol{\phi}}_k,
\end{equation}
where \(R_c(\phi_k)\) denotes the local loop radius and \(\hat{\boldsymbol{\phi}}_k\) is the tangential direction along the toroidal ring. The corresponding toroidal moment becomes
\begin{equation}
\mathbf{T}
=
\frac{I\pi R}{2}
\sum_{k=0}^{N-1}
R_c^2(\phi_k)\,
\hat{\mathbf z},
\end{equation}
or equivalently,
\begin{equation}
\mathbf{T}
=
\frac{N}{2}
I\pi R
\left\langle
R_c^2
\right\rangle
\hat{\mathbf z},
\end{equation}
where \(\langle R_c^2\rangle\) denotes the average squared loop radius.
To quantify a smooth lopsided distortion, we write
\begin{equation}
R_c(\phi)=R_0+\epsilon f(\phi),
\end{equation}
where \(R_0\) is the average loop radius, \(\epsilon\) characterizes the deformation strength, and \(f(\phi)\) describes its angular profile. For the representative symmetric deformation
\begin{equation}
f(\phi)=\cos\phi,
\end{equation}
the toroidal moment reduces to
\begin{equation}
\mathbf{T}
=
\frac{N}{2}
I\pi R
\left(
R_0^2+\frac{\epsilon^2}{2}
\right)
\hat{\mathbf z}.
\end{equation}
Thus, the toroidal moment remains aligned with the symmetry axis, while its magnitude is modified by the geometric distortion. In particular, symmetric lopsided deformations contribute only at second order in the deformation parameter, indicating that the toroidal state is robust against weak geometric asymmetries.
\subsubsection{Alternating radius}
\begin{figure}[H]
    \centering
    \includegraphics[width=0.95\linewidth]{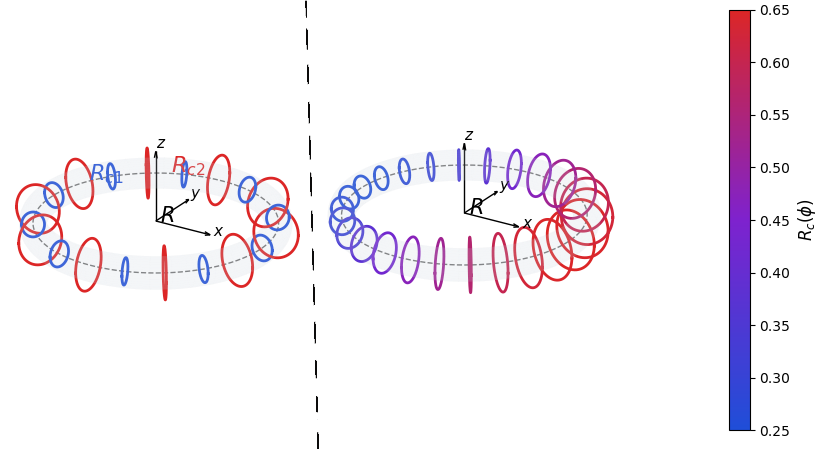}
    \caption{\textbf{Left:} Disteorted torus with radii $R_{c1}$ and $R_{c2}$
    and \textbf{Right:} lopsided toroidal geometry generated by the deformation \(R_c(\phi)=R_0+\epsilon\cos\phi\).
    }
    \label{fig:disttor}
\end{figure}

A simple special case is a ring composed of two types of loops (see Fig. ~\ref{fig:disttor}), with radii $R_{c1}$ and $R_{c2}$. Suppose a fraction $p$ of the $N$ loops have radius $R_{c1}$ and the remaining fraction $1-p$ have radius $R_{c2}$. Then
\begin{equation}
\sum_{k=0}^{N-1}R_{c,k}^2
=
Np\,R_{c1}^2 + N(1-p)\,R_{c2}^2.
\end{equation}
Thus, toroidal moment becomes,
\begin{equation}
\mathbf{T}
=
\frac{N I\pi R}{2}
\left(
p\,R_{c1}^2 + (1-p)\,R_{c2}^2
\right)\hat{\mathbf z}.
\label{eq:Ttworadii}
\end{equation}
This already captures how position-dependent loop size changes the toroidal response.\\
It is useful to distinguish the total magnetic dipole moment
\begin{equation}
\mathbf{m}_{\mathrm{tot}}
=
\sum_{k=0}^{N-1}\mathbf{m}_k
=
I\pi\sum_{k=0}^{N-1}R_{c,k}^2\,\hat{\boldsymbol\phi}_k
\label{eq:mtot}
\end{equation}
from the toroidal moment $\mathrm{T}$. The quantity $\mathbf{m}_{\mathrm{tot}}$ depends on both the magnitudes $R_{c,k}^2$ and the angular arrangement of the loops. If equal numbers of loops have radii $R_{c1}$ and $R_{c2}$, then $\mathbf{m}_{\mathrm{tot}}$ can vanish or not vanish depending on how the loops are distributed around the ring. If the two types are placed in a symmetric arrangement such that every loop has an opposite partner with the same radius, then pairwise cancellation occurs and $\mathbf{m}_{\mathrm{tot}}=0$. By contrast, an asymmetric arrangement may give $\mathbf{m}_{\mathrm{tot}}\neq 0$ even when the numbers are equal. If the numbers are unequal, one generically expects $\mathbf{m}_{\mathrm{tot}}\neq 0$.\\
The toroidal moment behaves differently. Since each term $\mathbf{r}_k\times \mathbf{m}_k$ points along the same $\hat{\mathbf z}$ direction provided the circulation sense is coherent, the toroidal moment does not suffer the same angular cancellation. Thus one may have
\begin{equation}
\mathbf{m}_{\mathrm{tot}}=0,
\qquad
\mathbf{T}\neq 0,
\end{equation}
which is precisely the characteristic toroidal situation.
\section{Interaction Between Two Discrete Toroidal Systems: The 4+4 Case}
\subsection{Nondeformed tori}
\begin{figure}[htbp]
    \centering
    \includegraphics[width=0.80\textwidth]{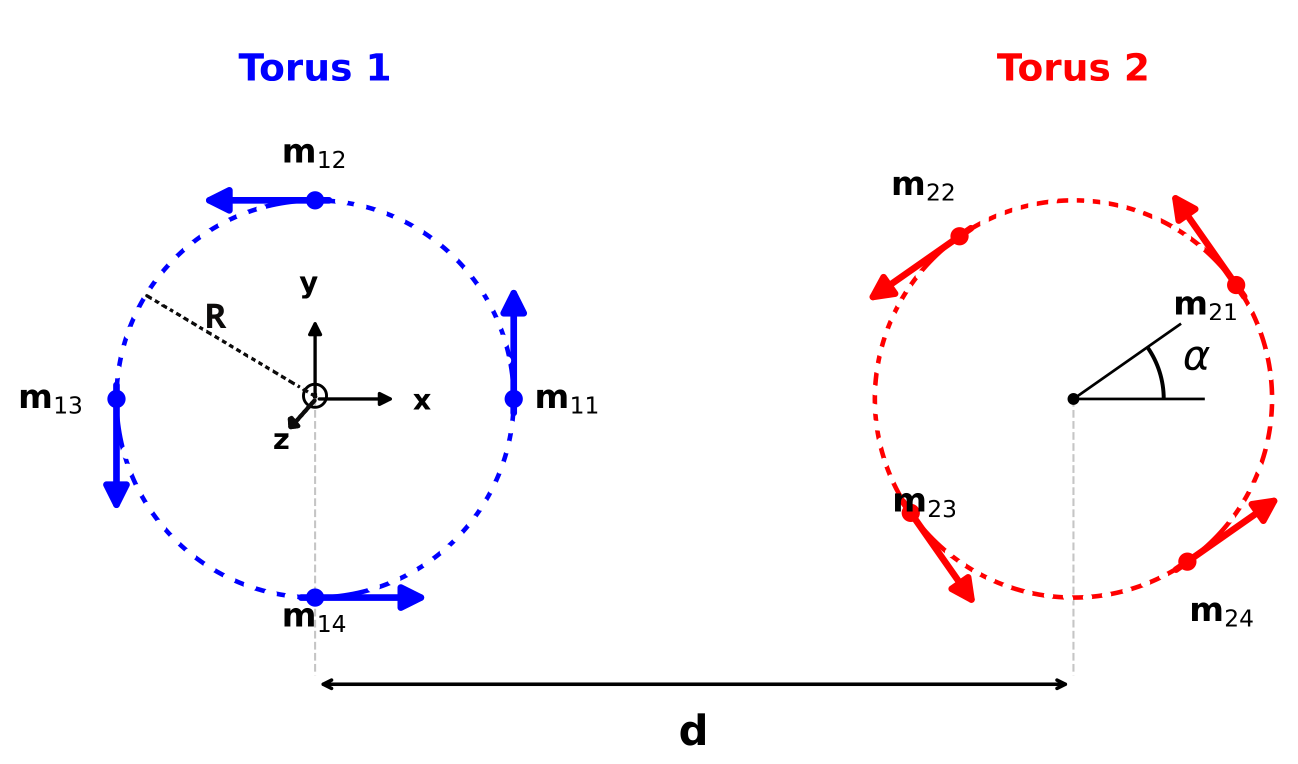}
    \caption{Schematic representation of the nondeformed discrete $4+4$
    spin-toroidal model. Each toroidal unit consists of four localized magnetic
    moments, $\mathbf{m}_{1i}$ and $\mathbf{m}_{2i}$ ($i=1,\ldots,4$), arranged
    tangentially on a circular ring of radius $R$. The center of Torus 1 defines
    the fixed Cartesian reference frame, while the center of Torus 2 is displaced
    by a distance $d$ along the $x$-direction. Torus 2 is rotated by an angle
    $\alpha$ about the common $\hat{z}$-axis, with both the moment positions and
    their orientations rotating together. The blue and red arrows denote the
    magnetic moments belonging to Torus 1 and Torus 2, respectively.}
    \label{fig:4plus4_schematic}
\end{figure}
To investigate the interaction between neighboring toroidal units, we consider two discrete spin-toroidal structures, each consisting of four localized magnetic moments arranged tangentially on a ring of radius \(R\). The centers of the two toroids are separated by a distance \(d\), and the second toroid may be rotated by an angle \(\alpha\) relative to the first. This minimal \(4+4\) configuration captures the essential geometric and magnetic features of interacting toroidal systems while remaining analytically tractable (see Fig. \ref{fig:4plus4_schematic}). The toroidal moment associated with each structure is given by
\begin{equation}
\mathbf T=
\frac{1}{2}
\sum_i
\mathbf r_i\times\mathbf m_i,
\end{equation}
where \(\mathbf r_i\) and \(\mathbf m_i\) denote the position and magnetic moment of the \(i\)-th loop, respectively. For the ideal configuration, the net magnetic dipole moment vanishes, while the toroidal moment remains finite and is directed along the symmetry axis.
The interaction between the two toroids is described within the magnetic dipole-dipole approximation (see supporting information for a detailed analysis). The interaction energy between moments \(k\) and \(\ell\) belonging to different toroids is
\begin{equation}
U_{k\ell}
=
\frac{\mu_0}{4\pi}
\left[
\frac{\mathbf m_{1k}\cdot\mathbf m_{2\ell}}
     {r_{k\ell}^{3}}
-
3
\frac{
(\mathbf m_{1k}\cdot\mathbf r_{k\ell})
(\mathbf m_{2\ell}\cdot\mathbf r_{k\ell})
}
{r_{k\ell}^{5}}
\right],
\end{equation}
where
\begin{equation}
\mathbf r_{k\ell}
=
\mathbf r_{2\ell}
-
\mathbf r_{1k}
\end{equation}
is the separation vector between the corresponding moments. The total interaction energy is obtained by summing over all pairwise contributions,
\begin{equation}
U_{\rm int}
=
\sum_{k=1}^{4}
\sum_{\ell=1}^{4}
U_{k\ell}.
\end{equation}
This discrete formulation provides a convenient framework for examining how \textcolor{black}{the dipole-dipole interaction between spin-toroidal configurations} is modified by geometric distortions, nonuniform moment distributions, and relative rotations between neighboring toroidal structures. In the following sections, we apply this formalism to ideal, elliptically deformed, and lopsided toroidal geometries and numerically analyze the resulting interaction landscapes.\\
{\color{black}
It is important to distinguish the collective toroidal character of each configuration from the microscopic interaction used here. The latter is entirely the conventional pairwise magnetic dipole-dipole interaction. The finite toroidal moment, together with the vanishing net magnetic dipole moment, characterizes the vortex-like texture, but does not by itself define an additional elementary ``toroidal-toroidal interaction''. A systematic discretization, Fourier, and far-field multipole analysis supporting this distinction is provided in the Supporting Information (Fig.s ~S3-S5).
}
\subsubsection{Numerical evaluation} \label{sec:nonDeformedNumericals}
\begin{figure}[H]
    \centering
    \includegraphics[width=0.7\textwidth]{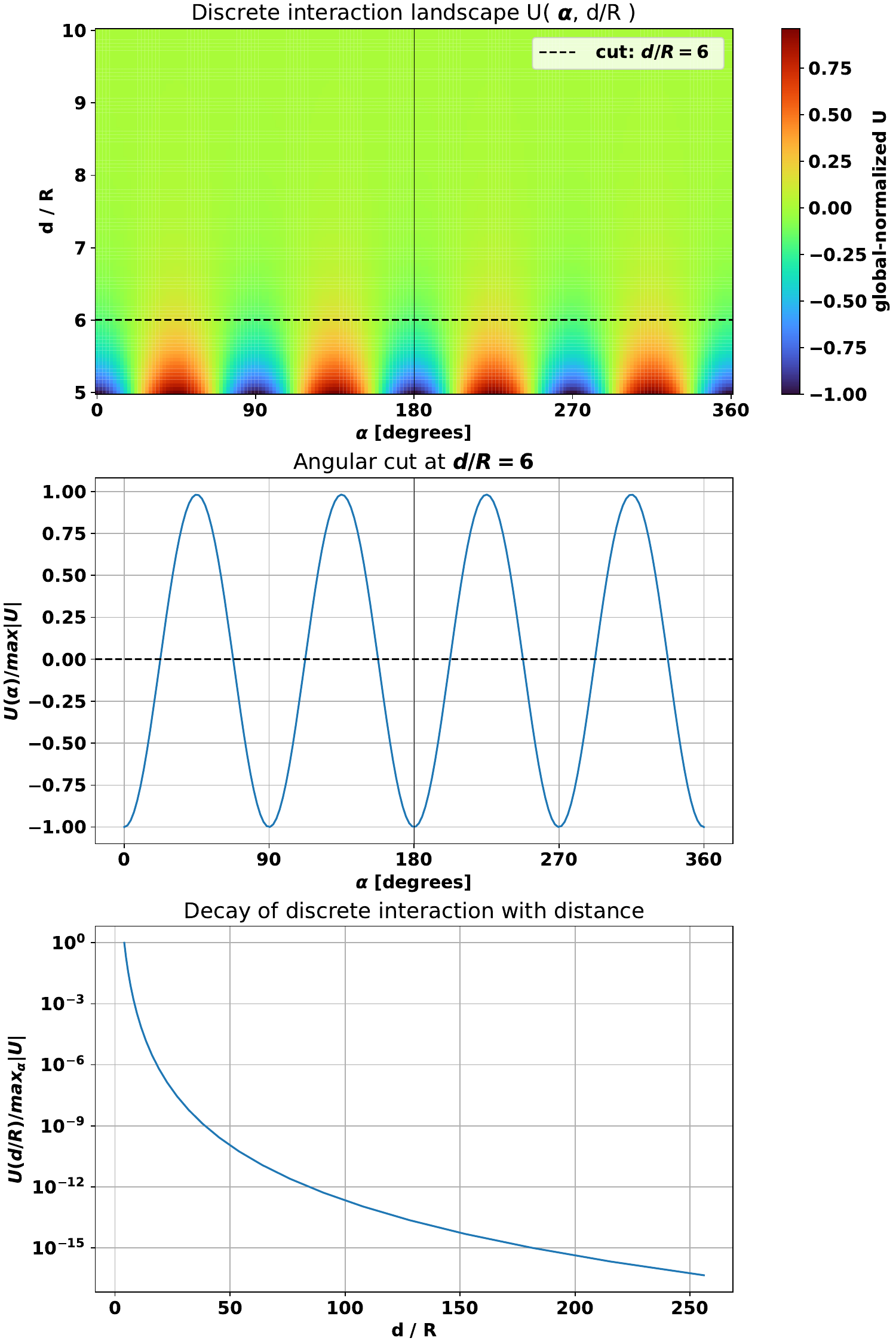}
    \caption{discrete interaction landscape $U(\alpha,d/R)$ for two non-deformed tori. 
    \textbf{Top:} Two-dimensional map of the interaction energy as a function of relative angle $\alpha$ 
    ( for $\mathbf{ \hat{ z } }$ rotation ) and normalized separation $d/R$, globally normalized to $[-1,1]$. \\
    \textbf{Middle:} Angular cut at $d/R=6$. \\
    \textbf{Bottom:} Decay of the interaction strength $\max_{\alpha}|U|$ with distance, plotted on a logarithmic scale.}
    \label{fig:discreteNondefLandscape}
\end{figure}

We first examine the interaction landscape of two identical undeformed toroidal structures within the discrete \(4+4\) model. All localized magnetic moments are taken to have equal magnitude,
\begin{equation}
m_{11}=m_{12}=m_{13}=m_{14}\equiv m_1,
\qquad
m_{21}=m_{22}=m_{23}=m_{24}\equiv m_2,
\end{equation}
corresponding to two ideal toroidal rings without geometric distortion. The interaction energy is evaluated from Eq.~(19) as a function of the relative rotation angle \(\alpha\) and the normalized separation \(d/R\).
The resulting interaction landscape is shown in Fig.~\ref{fig:discreteNondefLandscape} and S1.

The interaction landscape obtained from rotations about the toroidal symmetry axis \(\hat{\mathbf z}\) is shown in Fig.~\ref{fig:discreteNondefLandscape}. The discrete \((4+4)\) model exhibits a pronounced fourfold periodicity,
\begin{equation}
U(\alpha+90^\circ)=U(\alpha),
\end{equation}
reflecting the underlying symmetry of the four localized magnetic moments forming each torus. A rotation by \(90^\circ\) maps the toroidal structure onto itself, resulting in a repetition of the interaction landscape every quarter turn.\\
This behavior highlights the discrete nature of the model and demonstrates that the interaction energy is strongly influenced by the relative arrangement of the individual magnetic moments. Since rotations about \(\hat{\mathbf z}\) directly probe the internal toroidal structure and its symmetry, this rotational degree of freedom provides the most natural framework for analyzing the effects of geometric deformations. For this reason, the interaction landscapes discussed in the remainder of this work are evaluated using relative \(\hat{\mathbf z}\)-axis rotations. Changing the rotation axis from $\mathbf{ \hat{ z } }$ to $\mathbf{ \hat{ y } }$ produces a drastically different result as seen in Fig. S1.
To determine which features originate from the minimal $4+4$ discretization, we also studied ideal rings with $N=4,8,16,$ and $32$ tangential moments while keeping the total toroidal moment fixed. The leading Fourier harmonic is found at $n=N$, as required by the $C_N$ rotational symmetry. A complementary multipole analysis shows that the first non-vanishing external multipole has $\ell=N$, so that the asymptotic interaction of two identical ideal rings scales as
\begin{equation}
U_N(d)\propto d^{-(2N+1)}.
\end{equation}
High-precision numerical fits give exponents $9.0291$, $17.0359$, $33.0601$, and $65.0808$ for $N=4,8,16,$ and $32$, respectively, in excellent agreement with this selection rule. Thus, the pronounced angular landscape of the $4+4$ model should be understood as a property of the discrete internal magnetic structure rather than as a universal function of the toroidal moment alone. Full convergence, Fourier, and high-precision far-field results are given in the Supporting Information.



\subsection{Elliptically Deformed Toroidal Pair}

We now consider the interaction between two toroidal structures whose major rings are deformed into elliptical shapes (see Supporting Information and Fig. S2 for details). The deformation is characterized by the semi-axes \((R_a,R_b)\), which break the rotational symmetry of the circular torus discussed in the previous section. The magnetic moments remain tangential to the local ring geometry, while the interaction energy is evaluated using the dipole-dipole formalism introduced in Eq.~(17).

In contrast to the circular case, the elliptical deformation introduces a preferred geometric direction and consequently modifies both the angular dependence and the symmetry of the interaction landscape. The deformation affects the interaction through the altered dipole positions and the corresponding changes in the relative dipole orientations. As a result, the interaction can no longer be described solely by the symmetry classes of the undeformed geometry and must be evaluated numerically.

For the calculations shown below, both toroidal structures are taken to have identical semi-axes \(R_a\) and \(R_b\), with all local magnetic moments having equal magnitude.

\subsubsection{Numerical evaluation} \label{sec:ellipticNumericals}
We now evaluate the interaction landscape for two identical elliptically deformed toroidal structures. The deformation is characterized by the semi-axes \(R_a\) and \(R_b\), with the second torus rotated about the \(\hat{\mathbf z}\)-axis, i.e. along the direction of the toroidal moments. The interaction energy is computed using the pairwise dipole-dipole model introduced in Sec.~3.1.

\begin{figure}[H] 
    \centering
    \includegraphics[width=0.85\textwidth]{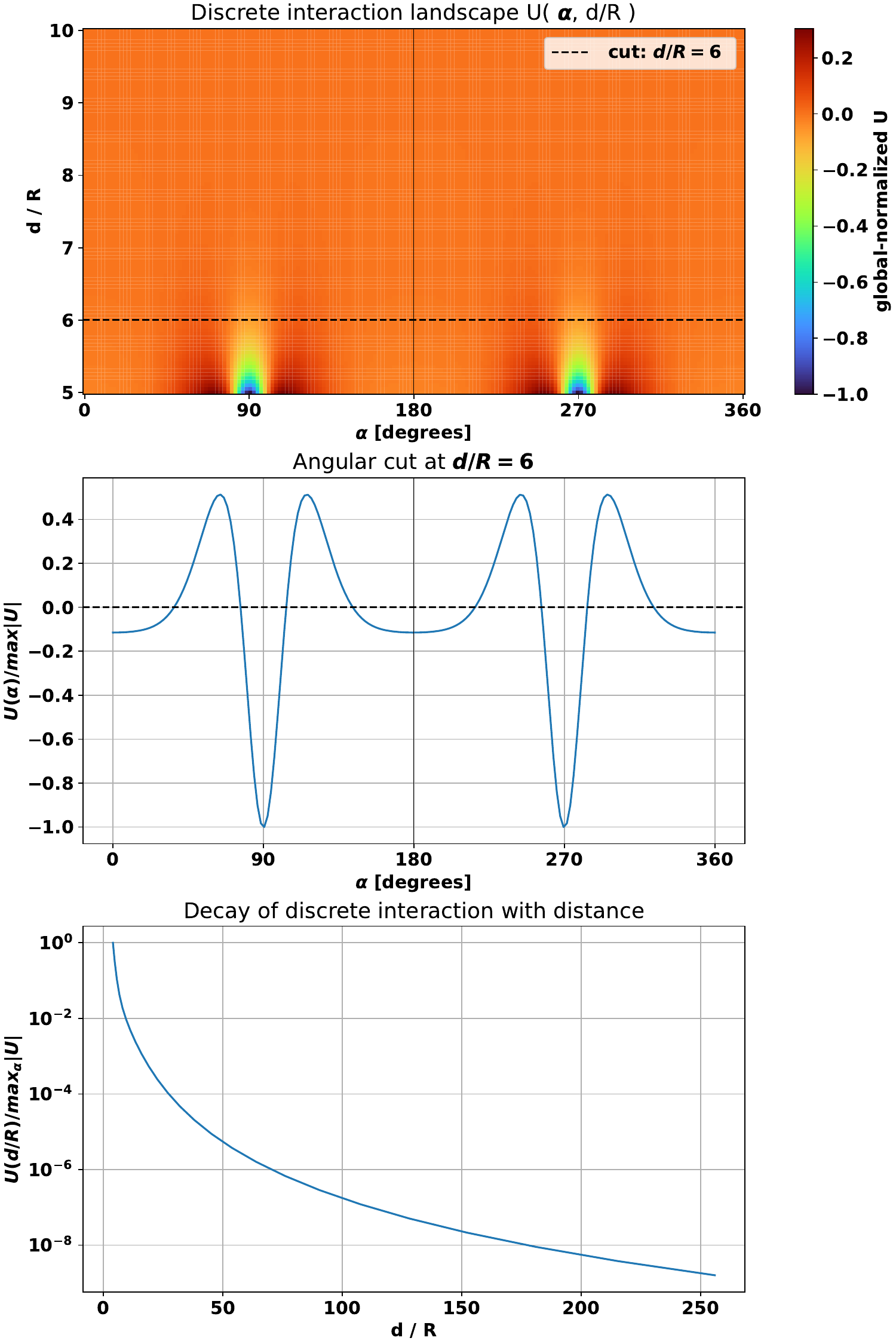}
    \caption{discrete interaction landscape $U(\alpha,d/R)$ for two elliptically deformed tori. 
    \textbf{Top:} Two-dimensional map of the interaction energy as a function of relative angle $\alpha$ 
    ( for $\mathbf{ \hat{ z } }$ rotation ) and normalized separation $d/R$, globally normalized to $[-1,1]$. \\
    \textbf{Middle:} Angular cut at $d/R=6$. \\ 
    \textbf{Bottom:} Decay of the interaction strength $\max_{\alpha}|U|$ with distance, plotted on a logarithmic scale.}
    \label{fig:discreteEllipticLandscape}
\end{figure}

For the results shown in Fig.~\ref{fig:discreteEllipticLandscape}, the two toroidal structures are taken to be identical, with

\begin{equation}
R_{1a}=R_{2a}=1,
\qquad
R_{1b}=R_{2b}=3,
\end{equation}

and all local magnetic moments chosen to have equal magnitude,

\begin{equation}
m_{1k}=m_{2\ell}=1,
\qquad
k,\ell \in \{1,2,3,4\}.
\end{equation}

Figure~\ref{fig:discreteEllipticLandscape} shows the resulting interaction energy as a function of the relative rotation angle \(\alpha\) and normalized separation \(d/R\). The interaction energies are normalized to emphasize the geometric dependence of the interaction landscape rather than their absolute magnitude.

 The interaction landscape for the elliptically deformed toroidal pair is shown in Fig.~\ref{fig:discreteEllipticLandscape}. In contrast to the undeformed case, the elliptical geometry introduces a pronounced anisotropy, leading to the appearance of deep local minima near \(\alpha \approx 90^\circ\) and \(270^\circ\). These minima arise when dipole pairs located near the vertices of the ellipses become nearly antiparallel while remaining in close proximity, resulting in strongly attractive dipole-dipole interactions that dominate the total \textcolor{black}{interaction} energy. The minima are accompanied by neighboring local maxima, which occur when the relative rotation moves these dipole pairs away from their optimal attractive configuration. The resulting change in dipole alignment increases the interaction energy and produces a characteristic minimum-maximum structure in the angular dependence of the interaction landscape.

A further consequence of the elliptical geometry is the emergence of a twofold rotational symmetry. The interaction energy remains invariant under the transformation

\begin{equation}
\alpha \rightarrow \alpha + 180^\circ ,
\end{equation}

which manifests itself as a periodic repetition of the energy landscape every half rotation of the second torus. This behavior reflects the residual symmetry of the elliptically deformed toroidal pair and highlights the strong influence of geometric anisotropy on \textcolor{black}{interactions between the discrete spin-toroidal configurations}. 
\subsection{Lopsided Toroidal Pair}

We next consider the interaction between two toroidal structures exhibiting the symmetric lopsided deformation introduced in Sec.~2.3. In this geometry, the local loop radius varies around the toroidal ring according to

\begin{equation}
R_c(\phi)=R_0+\epsilon \cos\phi,
\end{equation}

resulting in a non-uniform distribution of localized magnetic moments. Since the magnetic moment associated with each loop scales as \(m\propto R_c^2\), the geometric deformation can be incorporated naturally within the discrete model through a position-dependent set of magnetic moment magnitudes.

The interaction energy is evaluated using the dipole-dipole formalism introduced in Sec.~3.1 (see Supporting Information for details). In contrast to the undeformed and elliptically deformed cases, the lopsided geometry breaks the rotational symmetry of the toroidal ring and produces a strongly asymmetric distribution of local magnetic moments. Consequently, the interaction landscape becomes highly sensitive to the relative orientation of the two toroidal structures as well as to the deformation strength.
\subsubsection{Numerical evaluation}
\begin{figure}[H] 
    \centering
    \includegraphics[width=0.85\textwidth]{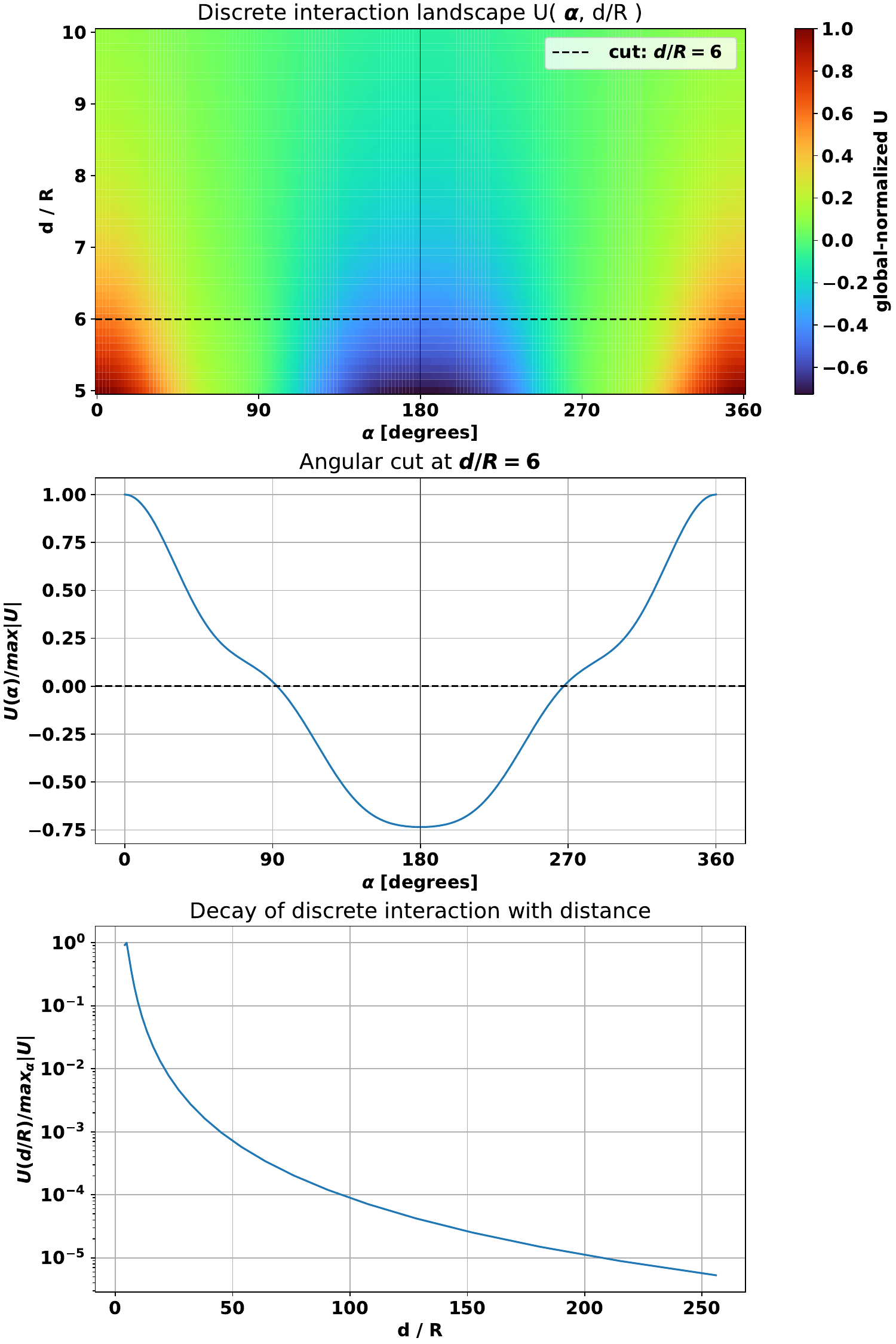}
    \caption{discrete interaction landscape $U(\alpha,d/R)$ for two symmetrically lopsided tori
    deformed to $\epsilon_{ rel } = - 0.3$ . 
    \textbf{Top:} Two-dimensional map of the interaction energy as a function of relative angle $\alpha$ 
    ( for $\mathbf{ \hat{ z } }$ rotation ) and normalized separation $d/R$, globally normalized to $[-1,1]$. \\
    \textbf{Middle:} Angular cut at $d/R=6$. \\
    \textbf{Bottom:} Decay of the interaction strength $\max_{\alpha}|U|$ with distance.}
    \label{fig:discreteGradualLandscape}
\end{figure}
To model the symmetric lopsided deformation, the magnetic moments are chosen according to the discrete representation of Eq.~(11). The undeformed torus corresponds to equal magnetic moments on all four sites, while the deformation introduces alternating larger and smaller moments reflecting the variation of the local loop radius around the ring.
For the results shown in Fig.~\ref{fig:discreteGradualLandscape}, both toroidal structures are taken to be identical, with deformation parameter \(\epsilon_{\rm rel}=\epsilon/R_0\). The interaction energy is evaluated as a function of the relative rotation angle \(\alpha\) and normalized separation \(d/R\).
The interaction landscape for the lopsided toroidal pair is shown in Fig.~\ref{fig:discreteGradualLandscape}. In contrast to the undeformed and elliptically deformed cases, the lopsided deformation breaks the higher rotational symmetries of the toroidal ring, leaving only the trivial invariance under a full rotation,
\begin{equation}
\alpha \rightarrow \alpha + 360^\circ.
\end{equation}
The resulting interaction energy therefore exhibits a strongly asymmetric angular dependence. A residual geometric symmetry remains due to the deformation profile \(R_c(\phi)=R_0+\epsilon\cos\phi\), which is invariant under reflection about the \(\phi=180^\circ\) axis.

For the deformation considered here (\(\epsilon_{\rm rel}<0\)), the largest magnetic moments are located at sites \(m_{13}\) and \(m_{23}\). Since the dipole-dipole interaction scales approximately as
\begin{equation}
|U_{k\ell}| \propto m_{1k}m_{2\ell},
\end{equation}
the interaction landscape is dominated by this pair. When the two tori are aligned (\(\alpha \approx 0^\circ\) or \(360^\circ\)), the dominant moments are arranged in a parallel configuration, producing a maximum in the interaction energy. Rotating one torus by \(180^\circ\) reverses their relative orientation and creates an antiparallel arrangement, resulting in the pronounced minimum observed near \(\alpha \approx 180^\circ\). The interaction strength decreases with increasing separation, but the characteristic asymmetry of the angular dependence remains a robust signature of the lopsided deformation.

\begin{figure}[H] 
    \raggedleft
    \includegraphics[width=0.9\textwidth]{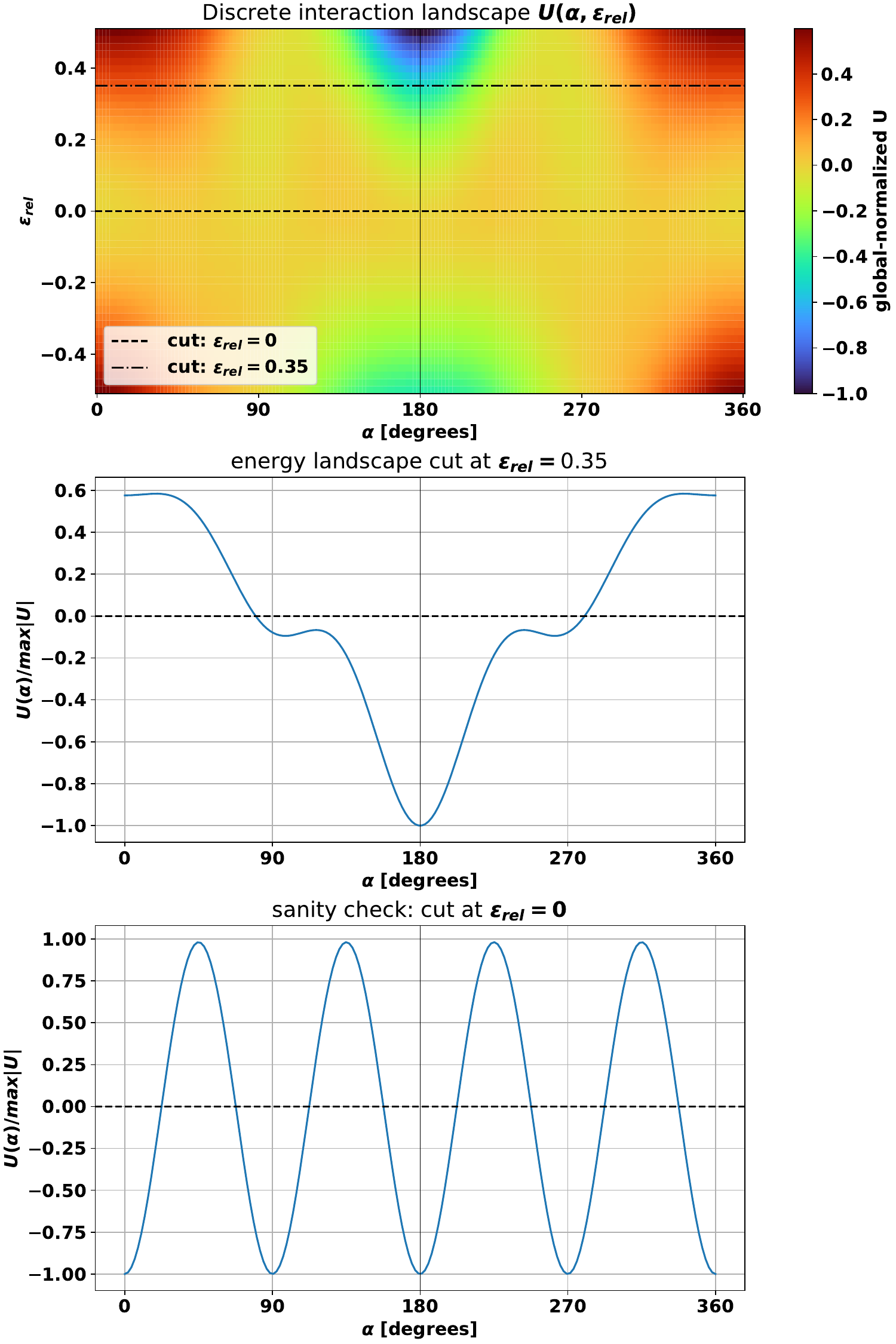}
    \caption{discrete interaction landscape $U(\alpha,\epsilon/R_0)$ for two symmetrically lopsided tori
    at central points distance $d = 6$.
    Two-dimensional map of the interaction energy as a function of the relative angle $\alpha$ 
    ( for $\mathbf{ \hat{ z } }$ rotation ) and deformation severity specified with $\epsilon_{ rel }$, globally normalized to $[-1,1]$.}
    \label{fig:varEpsilonLandscape}
\end{figure}

Another variant of the energy landscape shown in figure \ref{fig:varEpsilonLandscape}
presents \textcolor{black}{the pairwise interaction energy} depending on the deformation parameter $\epsilon_{ rel }$.
A sanity check for $\epsilon_{ rel } = 0$ recreates the shape of the $\alpha$-dependent interaction energy
of the ideal non-deformed case shown in figure \ref{fig:discreteNondefLandscape}.

At $\alpha \approx 180^\circ$ there is a remarkably deep ( see colorbar ), sharp interaction energy minimum
for large positive $\epsilon_{ rel }$ which contrasts with the much more gentle, wide minimum for
negative $\epsilon_{ rel }$ at the bottom of the landscape
( also visible in middle sections of fig. 
\ref{fig:discreteGradualLandscape} and \ref{fig:varEpsilonLandscape} ).

This difference arises from the pair $( \mathbf{m}_{11}, \mathbf{m}_{21} )$
now being the two largest magnetic moments out of the eight and thus influencing $U_\alpha$ the most
out of all the 16 intertoroidal pairings.

At $\alpha = 180^\circ$, the moments $( \mathbf{m}_{11}, \mathbf{m}_{21} )$ lay at a distance of only
$ d - 2R $ as opposed to the distance $ d + 2R $ of $( \mathbf{m}_{13}, \mathbf{m}_{23} )$
and since $|U_{kl}| \propto \frac{1}{r^3_{kl}}$,
the local minimum of $U_{11}$ for positive $\epsilon_{ rel }$ is much more pronounced compared to
that of $U_{33}$ for negative $\epsilon_{ rel }$ which carries over to the toroidal interaction energy
$U(\alpha) = \sum U_{kl}$.

\subsection{Toroidal Pair with Alternating Loop Radii}

Next, we consider a toroidal structure in which the local loop radii alternate between two values, as illustrated in Fig.~\ref{fig:disttor}. Since the magnetic moment associated with each loop scales as \(m\propto R_c^2\), the deformation can be represented within the discrete model by two alternating magnetic moment magnitudes. This geometry provides a simple realization of a non-uniform toroidal structure while preserving a higher degree of symmetry than the smoothly lopsided case discussed above.

The interaction energy is evaluated using the same dipole-dipole formalism introduced in Sec.~3.1. The alternating arrangement modifies the relative importance of individual dipole pairs and consequently alters the angular dependence of \textcolor{black}{interaction between the two discrete spin-toroidal configurations.}

\subsubsection{Numerical evaluation} \label{alternatingNumericals}

We specify both tori to be initially congruent, i.e.

\begin{align}
    m_{1c} &= m_{2c} = m_c \\
    m_{1d} &= m_{2d} = m_d
\end{align}

and again calculate the landscape of the \textcolor{black}{interaction energy} $U$ depending on their relative distance and orientation.

\begin{figure}[H]
    \centering
    \includegraphics[width=0.85\textwidth]{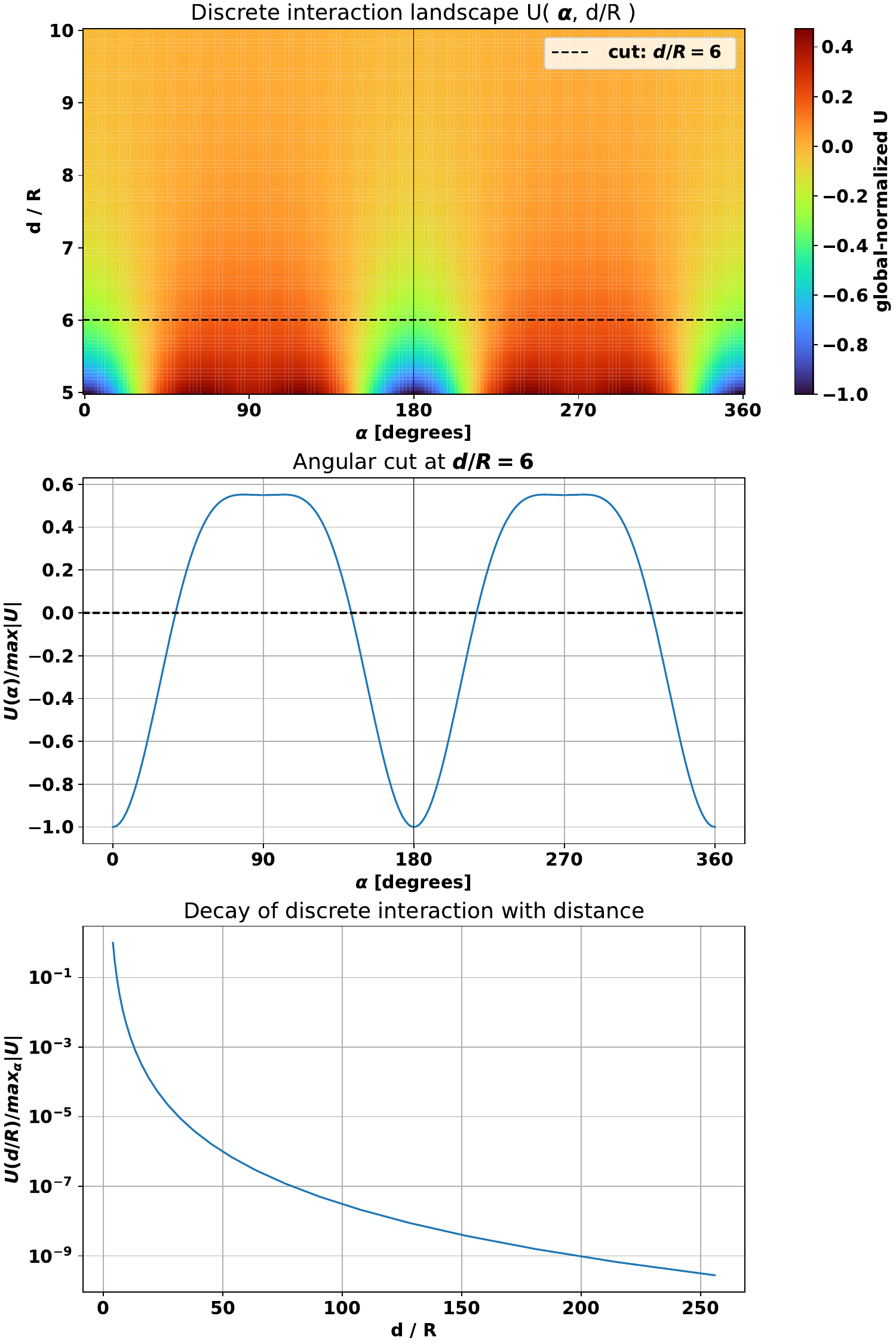}
    \caption{discrete interaction landscape $U(\alpha,d/R)$ for two tori with alternating magnetic moments
    ( $m_d > m_c$ ). 
    \textbf{Top:} Two-dimensional map of the interaction energy as a function of relative angle $\alpha$ 
    ( for $\mathbf{ \hat{ z } }$ rotation ) and normalized separation $d/R$, globally normalized to $[-1,1]$. \\
    \textbf{Middle:} Angular cut at $d/R=6$. \\
    \textbf{Bottom:} Decay of the interaction strength $\max_{\alpha}|U|$ with distance.}
    \label{fig:discreteAlternatingLandscape}
\end{figure}

The energy landscape fig. \ref{fig:discreteAlternatingLandscape} shows a periodicity of
$180^\circ$ in the $\alpha$-direction, i.e. $U$ is invariant under translations
$\alpha \rightarrow \alpha + 180^\circ$.
This angular translation symmetry stems from the definition of the alternating moments case
( see fig. \ref{fig:disttor}, left ) which gives the tori a rotational symmetry by the angular width of two placement intervals of the internal magnetic moments, i.e.

\begin{equation}
    \Delta\alpha = 360^\circ \cdot \frac{2}{N} = 180^\circ
\end{equation}

since $N=4$ for both tori.

Within both $\alpha$-periods of $U$ visible in fig. \ref{fig:discreteAlternatingLandscape},
a local maximum of $U$ lies at the center of the period and local minima at the borders.
In the first period their locations are

\begin{equation}
    \alpha_{U_{min}} \approx 0^\circ, \quad \alpha_{U_{max}} \approx 90^\circ.
\end{equation}

This results from the condition $m_d > m_c$ ( see Fig.\ref{fig:discreteAlternatingLandscape} ).
It causes pairs $( \mathbf{m_{1k}}, \mathbf{m_{2l}} )$ to have a particularly strong influence on the total interaction energy $U$ if both moments have absolute value $m_d$.
It also dictates that $( \mathbf{m}_{12}, \mathbf{m}_{22} )$ and $( \mathbf{m}_{14}, \mathbf{m}_{24} )$
are intertoroidal pairs carrying two of the larger magnetic moments $m_d$ each.
As the tori are congruent at $\alpha = 0^\circ$ both magnetic moments in each pair are oriented parallel along one line which is an arrangement of low $U_{kl}$ and consequently low $U(\alpha) $.

$\alpha \approx 90^\circ$ is situated approximately in the center between two local minima of $U$
( due to the $180^\circ$-periodicity ) so the aforementioned large magnetic moments are rotated
completely out of their parallel arrangement, maximizing $U(\alpha)$ as a sum of the $U_{kl}$.
\section{Tilt-Induced Anisotropy in Spin-Toroidal Rings}

In realistic molecular toroids, the local magnetic moments are not expected to remain perfectly tangential to the toroidal ring. Ligand-field anisotropy and local crystal environments may induce canting of the magnetic moments away from the ideal toroidal configuration. To investigate the consequences of such deviations, we consider two representative tilt scenarios: a uniform tilt of all moments and a pairwise tilt that preserves a vanishing net magnetic moment.

\subsection{Uniform Tilt} \label{sec:rectTilt}

We first consider a uniform tilt of all magnetic moments by an angle \(\theta\) away from the tangential direction. The toroidal and magnetic dipole moments become

\begin{equation}
T(\theta)=T_0\cos\theta,
\end{equation}

and

\begin{equation}
M(\theta)=M_0\sin\theta,
\end{equation}

respectively, where \(T_0\) denotes the toroidal moment of the ideal configuration and \(M_0=Nm\). Thus, increasing \(\theta\) continuously transforms a purely toroidal state into a conventional magnetic-dipolar state.

To describe the energetic cost of the tilt, we introduce the phenomenological anisotropy energy

\begin{equation}
E(\theta)=E_0+K\sin^2\theta,
\end{equation}

where \(K\) is an effective anisotropy constant. \rev{This term is a phenomenological modeling assumption and is not derived from the microscopic pairwise dipole-dipole Hamiltonian used for the interaction landscapes.} Figure~9 summarizes the evolution of the toroidal, magnetic, and energetic contributions as a function of the tilt angle.
\begin{figure}[H]
\centering
\includegraphics[width=0.85\textwidth]{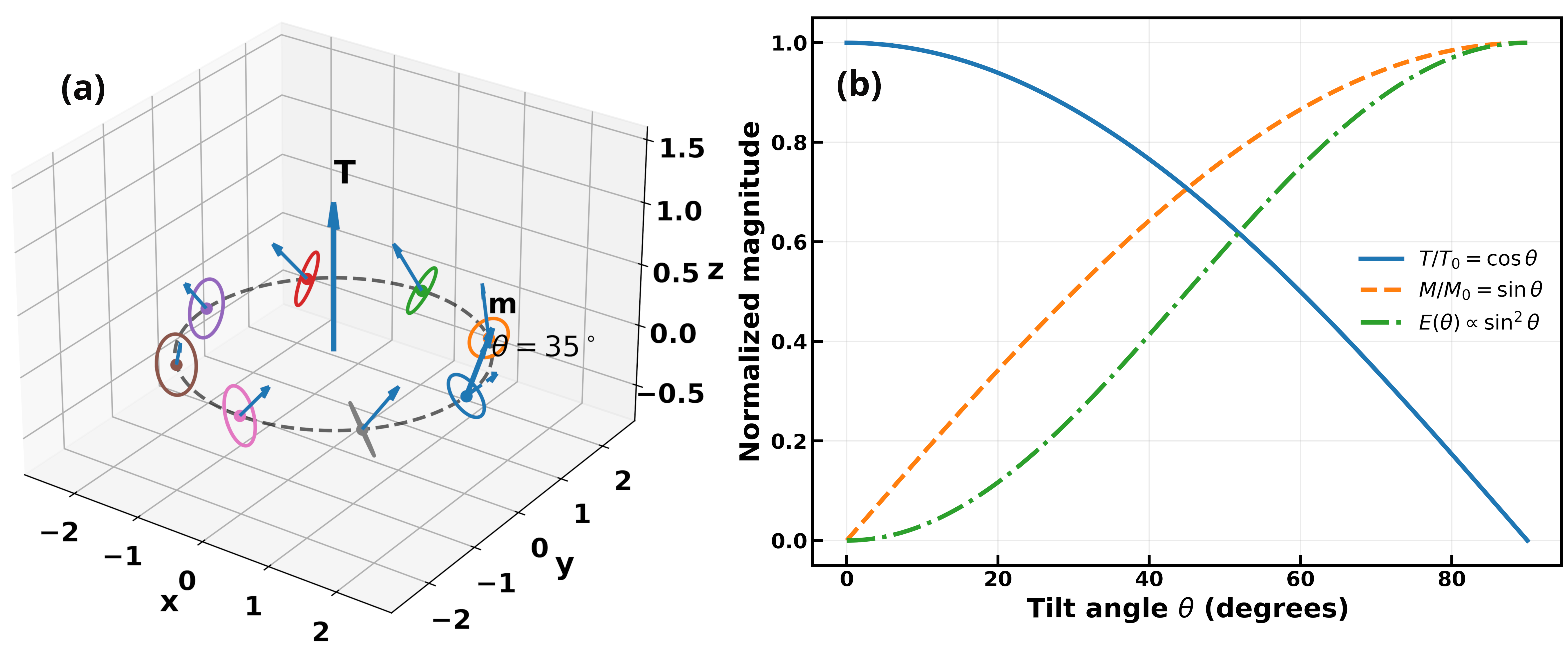}
\caption{(a) Schematic representation of tilted magnetic moments $\mathbf{m}$ arranged head-to-tail along a closed ring, generating a net toroidal moment $\mathbf{T}$ oriented along the symmetry axis. 
(b) Evolution of the normalized toroidal and magnetic components as a function of the tilt angle $\theta$. The toroidal moment scales as $T/T_0=\cos\theta$, the net magnetic component scales as $M/M_0=\sin\theta$, and the corresponding anisotropy energy follows $E(\theta)\propto\sin^2\theta$.
 }
\label{fig:tilt-ring}
\end{figure}

\subsubsection{Numerical $4+4$-case interaction}

We again numerically build an interaction energy landscape between two tori for which $N=4$ respectively, deviating from the non-deformed variant 
( for $\theta = 0^\circ$ ) as described in section \ref{sec:rectTilt}.
$\theta$ will be chosen between
$0^\circ$ and $90^\circ$
in order to capture both the toroidal and the introduced residual magnetic moment effects on the interaction energy.
Therefore it is optimal to fix the distance $d$ between the two center points
( along the $\hat{ \mathbf{ x }}$ axis )
and use the two available axes of the energy plot to map out a combined 
$\hat{ \mathbf{ y }}$ and $\hat{ \mathbf{ z }}$ axis rotation.

\begin{figure}[H] 
    \centering
    \includegraphics[width=0.85\textwidth]{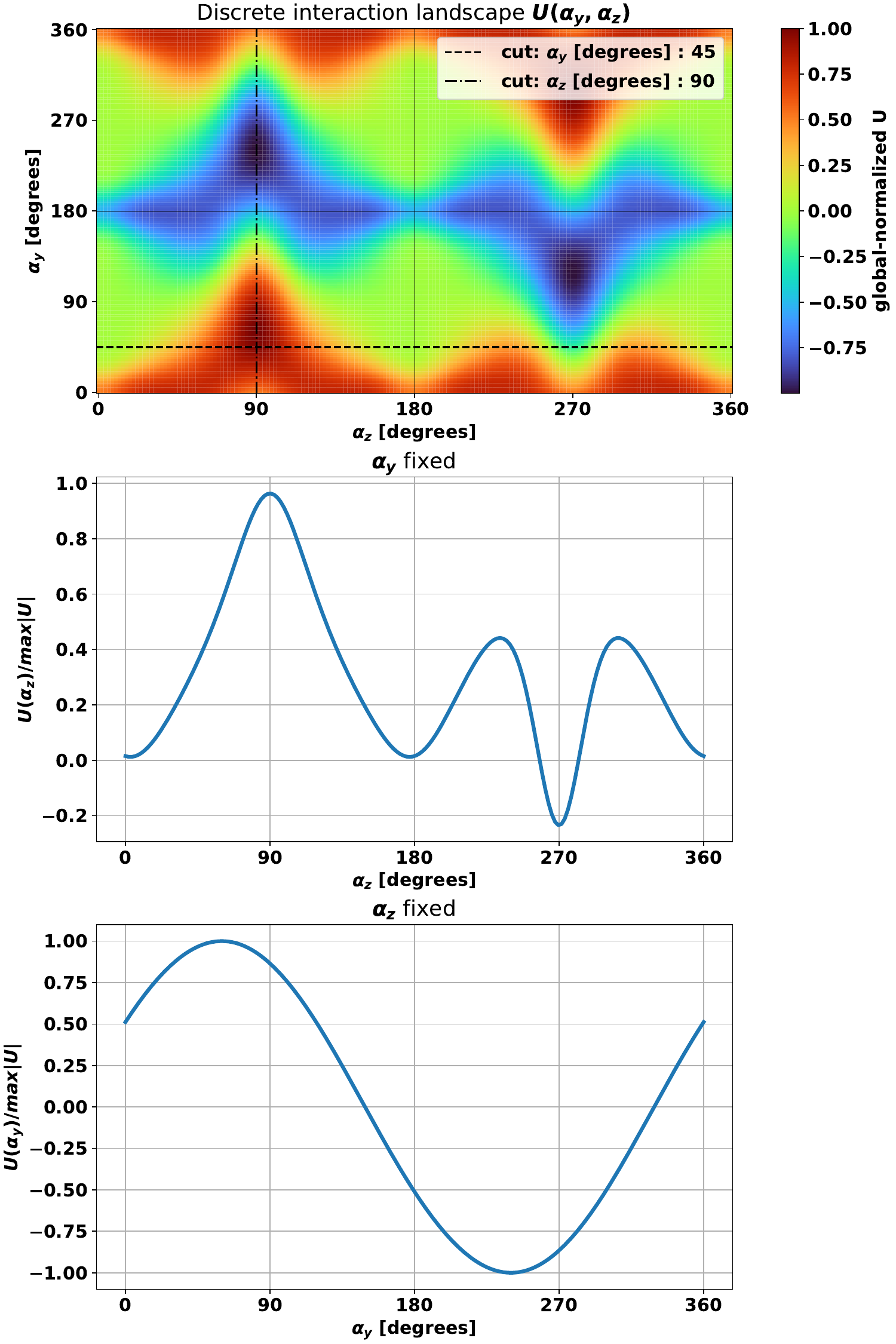}
    \caption{discrete interaction landscape $U( \alpha_y, \alpha_z )$ for two tori
    partially rectified at $\theta = 45^\circ$, $d / R = 3$. 
    \textbf{Top:} Two-dimensional map of the interaction energy as a function of relative angle $\alpha_y$ 
    ( for $\mathbf{ \hat{ y } }$ rotation )
    and relative angle $\alpha_z$ ( for $\mathbf{ \hat{ z } }$ rotation ),
    globally normalized to $[-1,1]$. \\
    \textbf{Middle:} Angular cut at $\alpha_y = 45^\circ$. \\
    \textbf{Bottom:} Angular cut at $\alpha_z = 90^\circ$.}
    \label{fig:rectTiltInteractionLandscape}
\end{figure}

Figure \ref{fig:rectTiltInteractionLandscape} reveals two characteristic symmetries of the interaction landscape. First, the interaction energy exhibits an approximate half-periodicity in the \(\alpha_y\) direction, originating from the residual magnetic moments generated by the uniform tilt. The corresponding dipolar contribution depends on the relative orientation of these residual moments and introduces an additional magnetic component to the interaction beyond the purely toroidal contribution. As a result, the angular dependence deviates slightly from a simple cosine form, reflecting the coexistence of toroidal and magnetic interactions.

A second symmetry is observed under the combined transformation

\begin{equation}
U(180^\circ-\alpha_y,\,180^\circ+\alpha_z)
=
U(\alpha_y,\alpha_z),
\end{equation}

which reflects the residual geometric symmetry of the tilted toroidal pair. Together, these features demonstrate that uniform tilting not only reduces the toroidal moment but also introduces additional anisotropic interaction channels through the emergence of a finite magnetic dipole component.

\subsection{Pairwise Tilt}

We next consider a constrained pairwise-tilt configuration in which diametrically opposite magnetic moments are tilted such that their vector contributions cancel. Specifically, pairs of opposite magnetic moments are tilted by the same angle $\alpha_k$ within the local radial-tangential plane. Owing to the imposed inversion relation between the corresponding moment vectors, the contributions from each pair cancel exactly, yielding
\begin{equation}
    \mathbf{m}_{\mathrm{tot}} = 0 ,
\end{equation}
for arbitrary values of the tilt angles $\alpha_k$ within this construction.

The toroidal moment is modified according to

\begin{equation}
\mathbf T
=
Rm
\sum_{k=1}^{N/2}
\cos(\alpha_k)\,
\hat{\mathbf z},
\end{equation}

where the \(\alpha_k\) represent the independent tilt angles associated with each pair of opposing magnetic moments. Within this constrained configuration, the pairwise tilt therefore reduces the magnitude of the toroidal moment while preserving the absence of a net magnetic dipole moment.

In the limiting case

\begin{equation}
\alpha_k = 0
\qquad
\forall k,
\end{equation}

the ideal toroidal configuration is recovered,

\begin{equation}
\mathbf T
=
\mathbf T_0
=
\frac{NRm}{2}\,
\hat{\mathbf z},
\qquad
\mathbf m_{\mathrm{tot}}
=
0.
\end{equation}
Thus, pairwise tilting provides a simple mechanism for tuning the toroidal order parameter without introducing an additional magnetic dipole contribution.
\subsubsection{Numerical Evaluation}
\begin{figure}[H] 
    \centering
    \includegraphics[width=0.85\textwidth]{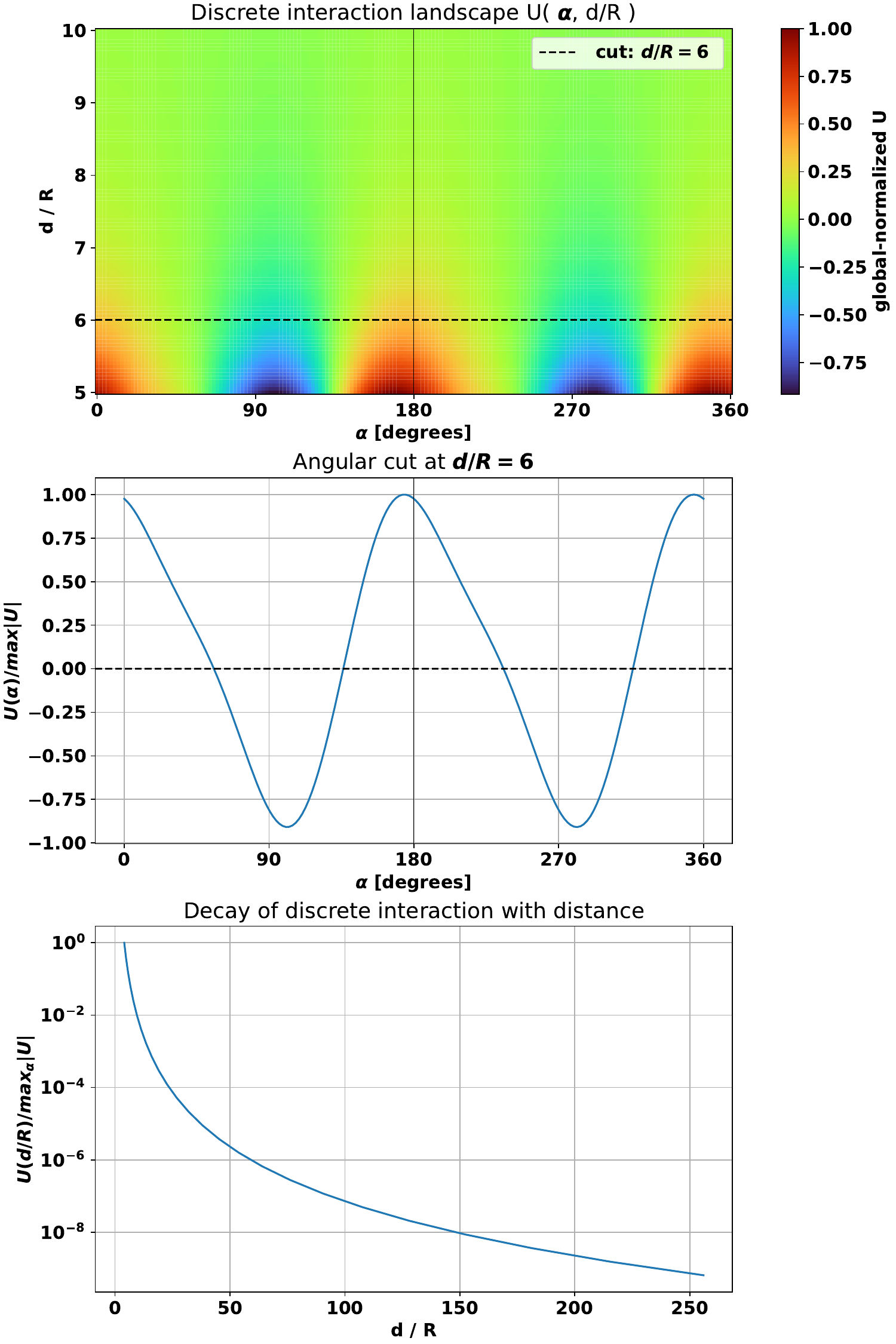}
    \caption {Discrete interaction landscape $U(\alpha,d/R)$ for two tori with the same internal pairwise tilt within the $\hat{ \mathbf{ x } } \hat{ \mathbf{ y } }$-plane by parameters $\alpha, \beta$. 
    \textbf{Top:} Two-dimensional map of the interaction energy
    at $\alpha = 40^\circ, \beta = -20^\circ$as a function of relative angle $\phi_z$ 
    ( for $\mathbf{ \hat{ z } }$ rotation ) and normalized separation $d/R$, 
    globally normalized to $[-1,1]$. 
    Middle: Angular cut at $d/R=6$;
    Bottom: Decay of the interaction strength $\max_{\alpha}|U|$ with distance.}
    \label{fig:pairwiseTiltLandscape}
\end{figure}
We consider two interacting toroidal structures in the \(4+4\) configuration, each containing four localized magnetic moments and separated by a distance \(d\). Both toroids are assumed to exhibit the same pairwise tilt pattern, characterized by two independent tilt angles,
\begin{equation}
\alpha_1=\alpha,
\qquad
\alpha_2=\beta.
\end{equation}
Unlike the uniform-tilt scenario discussed above, the constrained pairwise-tilt configuration considered here has a vanishing net magnetic moment by construction. The interaction landscape is therefore evaluated using only relative rotations about the \(\hat{\mathbf z}\)-axis, as in the undeformed and geometrically deformed toroidal configurations considered previously.
For the examples shown below, \(\alpha\) and \(\beta\) are chosen to have different magnitudes and opposite signs. This choice avoids the introduction of additional symmetries and provides a representative illustration of the general pairwise-tilt interaction scenario.
The interaction landscape for the pairwise-tilted toroidal pair is shown in Fig.~\ref{fig:pairwiseTiltLandscape}. Qualitatively, the resulting energy profile resembles that of the alternating-moment deformation discussed previously, reflecting the non-uniform distribution of effective magnetic moments around the toroidal ring. However, the extrema are shifted from the quarter-period positions observed in the alternating-moment case, and the interaction landscape becomes noticeably asymmetric.

This behavior originates from the radial tilting of the local magnetic moments. The tilt modifies the relative orientations of the dominant dipole pairs and therefore changes the rotational configurations that maximize or minimize the dipole-dipole interaction energy. As a result, the locations of the energy extrema are displaced from their positions in the undeformed and alternating-moment geometries.

Despite this asymmetry, the interaction energy retains a twofold rotational symmetry,

\begin{equation}
U(\phi_z+180^\circ)=U(\phi_z),
\end{equation}

which follows from the pairwise construction of the tilt pattern. Opposite magnetic moments remain equivalent under a half rotation of the toroidal structure, leading to the observed periodicity of the interaction landscape.

The pairwise-tilt scenario is particularly noteworthy because it \textcolor{black}{modifies the toroidal moment and the corresponding interaction landscape} without generating a net magnetic dipole moment. Consequently, interaction measurements may provide a means of detecting internal distortions of a toroidal spin texture even when conventional magnetic measurements indicate a vanishing total magnetic moment.

Taken together, these results show that structural and magnetic deformations leave characteristic symmetry signatures in the angular interaction landscape. The fourfold periodicity of the ideal four-site configuration reflects its underlying discrete rotational symmetry, whereas elliptical deformation reduces this symmetry and produces a predominantly twofold angular dependence. Lopsided distortions break the rotational symmetry more strongly, resulting in an asymmetric interaction landscape in which particular local magnetic-moment pairs become dominant. Changes in the magnitudes or orientations of the local moments similarly modify the angular dependence by altering the relative contributions of the interacting dipole pairs. These results demonstrate that the interaction landscape is governed not only by the overall toroidal moment, but also by the internal geometry and magnetic-moment arrangement of each toroidal unit. Consequently, toroidal structures with similar overall toroidal moments can exhibit qualitatively different orientation-dependent interactions.

\section{Molecular Tb$_4$ Toroidal Dimer as a Proof-of-Concept System}

To establish a qualitative connection between the idealized discrete toroidal interaction model and a realistic molecular system, we considered a tetranuclear Tb$_4$ molecular fragment inspired by experimentally reported single-molecule toroics\cite{wang2022ferroelectric}. In the experimental complexes, the Tb$^{3+}$ ions possess strong local magnetic anisotropy and arrange in a vortex-like configuration, producing a finite toroidal moment while maintaining a nearly vanishing net magnetic dipole moment. Motivated by this geometry, we extracted an isolated Tb$_4$ molecular unit from the crystallographic structure and used it as a simplified molecular realization of the discrete toroidal model.
In molecular toroics, the formation and stabilization of the individual toroidal magnetic texture is governed by microscopic interactions such as exchange coupling and local magnetic anisotropy\cite{soncini2008toroidal}, whereas the present model focuses on the subsequent, weaker dipolar interaction between already established toroidal units.

Instead of using the full periodic crystal environment, only a single molecular fragment was retained. This choice was made to isolate the intrinsic intermolecular orientation-dependent interaction between two toroidal molecular units while avoiding additional complications arising from long-range crystal packing, solvent molecules, and periodic-image interactions. The extracted molecular fragment preserves the essential Tb$_4$ magnetic core responsible for the toroidal arrangement. The present two-unit calculation is intended to characterize the local orientation-dependent interaction between a pair of toroidal molecular units rather than the total interaction energy or ground-state arrangement of an extended crystal. In a molecular crystal, several neighboring units may occur at comparable distances and can therefore contribute significantly to the total interaction.

\begin{figure}[H]
\centering
\includegraphics[width=0.85\textwidth]{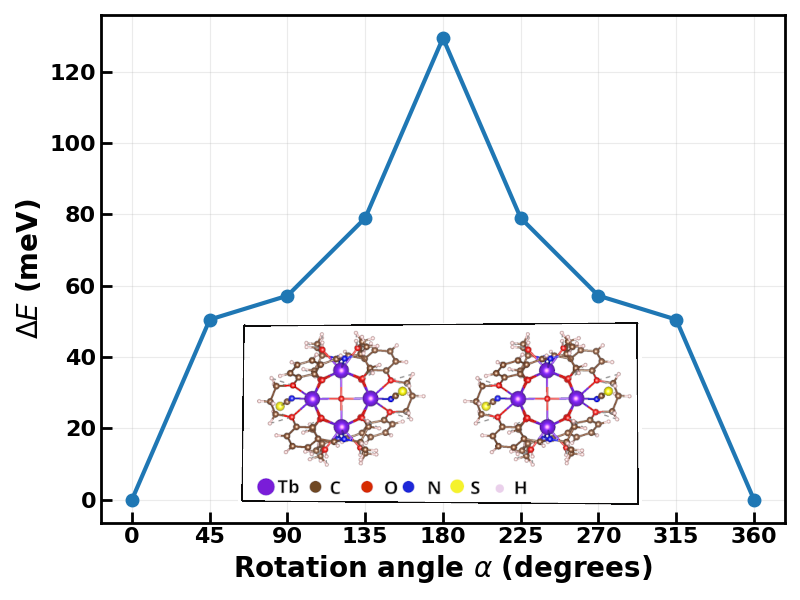}
\caption{
Orientation-dependent interaction energy landscape for two Tb$_4$ molecular toroidal units obtained from noncollinear spin-polarized DFT calculations including spin-orbit coupling, inset: Geometry of the Tb$_4$ molecular toroidal dimer used in the orientation-dependent interaction calculations.
}
\label{fig:tb4-E}
\end{figure}

Two such molecular units were then placed side by side at a fixed center-to-center separation distance $d$ (18~\AA, see Fig. \ref{fig:tb4-E}, and Fig. ~S6), while one molecule was rotated by an angle $\alpha$ with respect to the other. The calculations were performed using noncollinear spin-polarized density functional theory including spin-orbit coupling (SOC) in VASP.\cite{kresse1993ab, kresse1996efficiency, kresse1996efficient, kresse1999ultrasoft} The local Tb moments were initialized in a vortex-like arrangement corresponding to tangential magnetic directions around the Tb$_4$ core,
which mimics the circulating toroidal configuration discussed in the experimental literature\cite{wang2022ferroelectric}. \rev{The complete computational setup, including the exchange-correlation functional, treatment of the Tb $4f$ states, PAW datasets, plane-wave cutoff, convergence criteria, dispersion treatment, charge state, supercell/vacuum conditions, structural constraints, and magnetic initialization/convergence checks, is given in the Supporting Information.}

Single-point relative energies were calculated for several relative orientations:
\[
\alpha = 0^\circ,\ 45^\circ,\ 90^\circ,\ 135^\circ,\ 180^\circ.
\]

\[
\Delta E(\alpha)=E(\alpha)-E(0^\circ),
\]
which gives
\begin{align}
\Delta E(45^\circ)  &= 50.5~\mathrm{meV}, \\
\Delta E(90^\circ)  &= 57.2~\mathrm{meV}, \\
\Delta E(135^\circ) &= 79.0~\mathrm{meV}, \\
\Delta E(180^\circ) &= 129.5~\mathrm{meV}.
\end{align}

The calculations reveal a clear orientation-dependent interaction landscape,
with the aligned configuration ($\alpha = 0^\circ$) being energetically most
favorable, while progressively rotated configurations become increasingly less
stable (see Fig. \ref{fig:tb4-E}). The variation is smooth and monotonic over the angular range considered. However, these orientation-dependent energy differences represent changes in the total DFT energy of the molecular dimer and should not be interpreted as a direct measure of the toroidal magnetic interaction energy.

It is important to emphasize, that the total DFT energy contains
contributions from multiple sources,
\[
E_{\mathrm{total}}
=
E_{\mathrm{magnetic}}
+
E_{\mathrm{electrostatic}}
+
E_{\mathrm{dispersion}}
+
E_{\mathrm{steric}}
+
E_{\mathrm{ligand-ligand}}
+\cdots,
\]
and therefore does not represent a purely toroidal magnetic interaction energy.
At the chosen center-to-center separation of approximately $18~\mathrm{\AA}$,
direct orbital overlap and steric contact between the two molecular fragments
are expected to be substantially reduced. Consequently, the remaining
orientation-dependent energy is more likely to reflect long-range anisotropic
interactions, including electrostatic, dispersion, magnetic dipolar, and
spin-toroidal contributions.

Nevertheless, the total DFT energy cannot be uniquely decomposed into an isolated toroidal magnetic term. The calculations should therefore be interpreted
as a realistic molecular interaction landscape that is qualitatively consistent with the discrete spin-toroidal model, rather than as a direct quantitative
extraction of a pure toroidal coupling energy.

The discrete spin-toroidal model developed in the previous sections isolates
the geometric and magnetic contributions associated with localized circulating
spin textures, whereas the molecular calculations incorporate the full electronic
and chemical environment. The qualitative consistency between the two
approaches suggests that geometrically induced anisotropic interaction
landscapes may persist beyond the idealized discrete spin-toroidal description.

\section{Conclusions}
In this work, we developed a discrete spin-toroidal framework to analyze the effects of geometric distortions and spin reorientation on toroidal moments and toroidal interaction landscapes.  We considered an in-plane tangential easy axes of the spins resulting in toroidal moment remaining aligned with the symmetry axis while the net magnetic dipole moment vanishes. Introducing geometric distortions revealed that the toroidal response is controlled primarily by the spatial distribution of the local magnetic moments but the orientation remains aligned with the symmetry axis. We considered the interaction landscape for two such toroidal rings and found that the interaction pattern is controlled primarily by the relative orientation and strength of the nearest and next nearest neighbour local magnetic moments. In particular, elliptical and lopsided deformations modify the interaction landscape by breaking rotational symmetries and introducing anisotropic angular dependencies. 

The pairwise dipole formulation enabled a detailed analysis of \rev{the interaction between discrete spin-toroidal textures}. For undeformed structures, the interaction energy exhibits well-defined angular periodicities determined by the symmetry of the discrete magnetic-moment arrangement. Geometric distortions reduce these symmetries and generate new interaction minima and maxima, demonstrating that the interaction landscape is highly sensitive to the internal structure of the toroidal units. \rev{For an ideal $N$-site ring, the $C_N$ selection rule gives a leading angular harmonic $n=N$ and a first non-zero external multipole $\ell=N$, resulting in the asymptotic scaling $U_N(d)\propto d^{-(2N+1)}$. This explicitly shows that the far-field interaction is governed by the discrete higher-multipole content and not by the toroidal moment alone.}

We further examined the role of spin tilting. Uniform tilt of all magnetic moments continuously transforms a purely toroidal configuration into a magnetic-dipolar one, with the toroidal and magnetic components varying as $T/T_0=\cos\theta$ and $M/M_0=\sin\theta$, respectively. In contrast, a constrained pairwise tilting preserves a vanishing net magnetic moment while still producing measurable modifications of the interaction landscape. An important outcome of this analysis is that the toroidal moment alone does not fully determine the interaction between two toroidal units, since the internal geometry and spatial arrangement of the constituent magnetic moments can substantially modify the angular interaction landscape

Finally, a proof-of-concept study based on a Tb$_4$ molecular toroidal dimer demonstrated that realistic molecular systems exhibit orientation-dependent interaction energies qualitatively consistent with the trends predicted by the discrete spin-toroidal model. Although the total density-functional-theory energies contain electrostatic, steric, and electronic contributions beyond pure toroidal coupling, \rev{the calculation is used only to establish a non-trivial molecular orientation-dependent total-energy landscape and not as a quantitative extraction of a pure toroidal magnetic interaction.}

 Thus, the framework developed here offers a simple and physically transparent platform for studying spin-toroidal matter and may be extended to larger spin networks, molecular toroics, and magnetoelectric systems where toroidal order plays a central role.

\section*{Author contributions}

 Numerical simulations, analysis, and tests  were done by AB and HP. All authors contributed to the  data interpretation, writing, and editing of the manuscript. 

\section*{Acknowledgments}
The work has been supported by the DFG under project nr. 465098690.

\section*{Financial disclosure}

None reported.

\section*{Conflict of interest}

The authors declare no potential conflict of interests.

\section*{Data Availability Statement}

The data beyond those reproducible from the equation in the text are available from the  authors upon reasonable request.
\newpage
\appendix
\textbf{\Large \underline{Appendices}}:\\

\section{Single undeformed torus}
\subsection{Explicit component calculation}

For completeness, we verify the result in Cartesian coordinates. The position and magnetic moment vectors are

\begin{equation}
	\mathbf{r}_k
	=
	(R\cos\phi_k,\,R\sin\phi_k,\,0),
\end{equation}

and

\begin{equation}
	\mathbf{m}_k
	=
	I\pi R_c^2
	(-\sin\phi_k,\,\cos\phi_k,\,0),
\end{equation}

respectively.

Therefore,
\begin{align}
	\mathbf{r}_k\times \mathbf{m}_k
	&=
	\begin{vmatrix}
		\hat{\mathbf x} & \hat{\mathbf y} & \hat{\mathbf z}\\
		R\cos\phi_k & R\sin\phi_k & 0\\
		-I\pi R_c^2\sin\phi_k & I\pi R_c^2\cos\phi_k & 0
	\end{vmatrix}
	\nonumber\\[4pt]
	&=
	\hat{\mathbf z}
	\left[
	R\cos\phi_k\,(I\pi R_c^2\cos\phi_k)
	-
	R\sin\phi_k\,(-I\pi R_c^2\sin\phi_k)
	\right]
	\nonumber\\[4pt]
	&=
	\hat{\mathbf z}\,
	R I\pi R_c^2(\cos^2\phi_k+\sin^2\phi_k)
	\nonumber\\[4pt]
	&=
	R I\pi R_c^2\,\hat{\mathbf z}.
\end{align}
Again, this yields

\begin{equation}
	\mathbf{T}
	=
	\frac{N}{2}
	R I\pi R_c^2
	\hat{\mathbf z},
\end{equation}

in agreement with the result presented in the main text.

\subsection{Condition on circulation sense}

The result above assumes that all loops have the same circulation sense, so that all magnetic moments point along $+\hat{\boldsymbol\phi}_k$ or all along $-\hat{\boldsymbol\phi}_k$. This is essential. If neighboring loops have inconsistent circulation directions, then some $\mathbf{m}_k$ reverse sign and the toroidal contributions can cancel. Hence a nontrivial toroidal state requires coherent circulation.

If all loops are reversed simultaneously, then
\begin{equation}
	\mathbf{m}_k \to -\mathbf{m}_k,
	\qquad
	\mathbf{T}\to -\mathbf{T}.
\end{equation}
Thus the toroidal moment changes sign under global reversal of current circulation, as expected.

\subsection{Order parameter description}

To model imperfect toroidal order, we introduce an order parameter $p_k$ for each loop and write
\begin{equation}
	\mathbf{m}_k
	=
	p_k\,m_0\,\hat{\boldsymbol\phi}_k,
	\qquad
	m_0 = I\pi R_c^2.
	\label{eq:pkdef}
\end{equation}
Then
\begin{equation}
	\mathbf{T}
	=
	\frac{1}{2}\sum_{k=0}^{N-1}\mathbf{r}_k\times \mathbf{m}_k
	=
	\frac{1}{2}\sum_{k=0}^{N-1}
	R\,p_k\,m_0\,\hat{\mathbf z}.
\end{equation}
Hence
\begin{equation}
	\mathbf{T}
	=
	\frac{R m_0}{2}
	\left(\sum_{k=0}^{N-1}p_k\right)\hat{\mathbf z}.
	\label{eq:Tpk}
\end{equation}
If all loops have the same degree of order, $p_k=p$, then
\begin{equation}
	\mathbf{T}
	=
	p\,\mathbf{T}_0,
	\qquad
	\mathbf{T}_0=\frac{N}{2}I\pi R R_c^2\,\hat{\mathbf z}.
	\label{eq:Tglobalp}
\end{equation}
Thus
\begin{equation}
	p=1 \;\Rightarrow\; \text{perfect toroidal order},
	\qquad
	p=0 \;\Rightarrow\; \text{no toroidal moment}.
\end{equation}
\subsection{Deformation I: elliptical ring of loop centers}

We now deform the major ring from a circle to an ellipse. The center of the $k$-th loop is taken as
\begin{equation}
	\mathbf{r}_k
	=
	(R_x\cos\phi_k,\,
	R_y\sin\phi_k,\,
	0).
	\label{eq:ellipsecenters}
\end{equation}
If we still choose the loop magnetic moments tangential to the ring, then the tangent vector is
\begin{equation}
	\frac{d\mathbf{r}}{d\phi}
	=
	(-R_x\sin\phi,\,
	R_y\cos\phi,\,
	0),
\end{equation}
whose norm is
\begin{equation}
	\lambda(\phi)
	=
	\sqrt{R_x^2\sin^2\phi + R_y^2\cos^2\phi}.
\end{equation}
Therefore the unit tangent is
\begin{equation}
	\hat{\mathbf t}(\phi)
	=
	\frac{1}{\lambda(\phi)}
	(-R_x\sin\phi,\,
	R_y\cos\phi,\,
	0),
\end{equation}
and the loop moment becomes
\begin{equation}
	\mathbf{m}(\phi)
	=
	I\pi R_c^2\,\hat{\mathbf t}(\phi).
\end{equation}
Then
\begin{align}
	\mathbf{r}(\phi)\times \mathbf{m}(\phi)
	&=
	\begin{vmatrix}
		\hat{\mathbf x}&\hat{\mathbf y}&\hat{\mathbf z}\\
		R_x\cos\phi&R_y\sin\phi&0\\
		-\dfrac{I\pi R_c^2 R_x\sin\phi}{\lambda(\phi)}
		&
		\dfrac{I\pi R_c^2 R_y\cos\phi}{\lambda(\phi)}
		&0
	\end{vmatrix}
	\nonumber\\[4pt]
	&=
	\hat{\mathbf z}\,
	\frac{I\pi R_c^2}{\lambda(\phi)}
	\left(
	R_xR_y\cos^2\phi + R_xR_y\sin^2\phi
	\right)
	\nonumber\\[4pt]
	&=
	\frac{I\pi R_c^2 R_xR_y}{\lambda(\phi)}\,\hat{\mathbf z}.
\end{align}
Hence
\begin{equation}
	\mathbf{T}
	=
	\frac{1}{2}
	\sum_{k=0}^{N-1}
	\frac{I\pi R_c^2 R_xR_y}
	{\sqrt{R_x^2\sin^2\phi_k+R_y^2\cos^2\phi_k}}
	\,\hat{\mathbf z}.
	\label{eq:Tellipse}
\end{equation}
Thus the toroidal moment remains along $\hat{\mathbf z}$, but its magnitude is modified by the breaking of rotational symmetry.
\section{Interaction Between Two Discrete Toroidal Systems: The \(4+4\) Case} 
\label{sec:4+4_Introduction}

We consider two discrete toroidal systems, each composed of four localized magnetic dipoles arranged on a ring. The first torus has radius \(R_1\), the second torus has radius \(R_2\), and their centers are separated by a distance \(d\) along the \(\hat y\)-direction. The second torus is rotated by an angle \(\alpha\) about the \(\hat x\)-axis.

For convenience, we define
\begin{equation}
	c \equiv \cos\alpha, \qquad s \equiv \sin\alpha .
\end{equation}

\subsection{Dipole directions and positions} \label{sec:4+4_nonDeformedSpecs}

For torus 1, the dipole directions are taken as
\begin{align}
	\hat{\mathbf m}_{11} &= -\hat x, & \\
	\hat{\mathbf m}_{12} &= -\hat y, & \\
	\hat{\mathbf m}_{13} &= \hat x, & \\
	\hat{\mathbf m}_{14} &= \hat y,
\end{align}
with corresponding position vectors
\begin{align}
	\mathbf r_{m_{11}} &= R_1 \hat y, & \\
	\mathbf r_{m_{12}} &= -R_1 \hat x, & \\
	\mathbf r_{m_{13}} &= -R_1 \hat y, & \\
	\mathbf r_{m_{14}} &= R_1 \hat x.
\end{align}

For torus 2, after rotation by angle \(\alpha\) about the \(\hat x\)-axis, the dipole directions become
\begin{align}
	\hat{\mathbf m}_{21} &= -\hat x, \\
	\hat{\mathbf m}_{22} &= -c\,\hat y + s\,\hat z, \\
	\hat{\mathbf m}_{23} &= \hat x, \\
	\hat{\mathbf m}_{24} &= c\,\hat y - s\,\hat z,
\end{align}
and the corresponding position vectors are
\begin{align}
	\mathbf r_{m_{21}} &= (d+R_2 c)\hat y - R_2 s\,\hat z, \\
	\mathbf r_{m_{22}} &= -R_2 \hat x + d \hat y, \\
	\mathbf r_{m_{23}} &= (d-R_2 c)\hat y + R_2 s\,\hat z, \\
	\mathbf r_{m_{24}} &= R_2 \hat x + d \hat y.
\end{align}

\subsection{Toroidal moments of the two tori}

For each torus, the discrete toroidal moment is
\begin{equation}
	\mathbf T = \frac12 \sum_{j=1}^{4} \mathbf r_j \times \mathbf m_j .
\end{equation}
Hence,
\begin{align}
	\mathbf T_1 &= \frac12 \sum_{k=1}^{4} \mathbf r_{m_{1k}} \times \mathbf m_{1k}, \\
	\mathbf T_2 &= \frac12 \sum_{\ell=1}^{4} \mathbf r_{m_{2\ell}} \times \mathbf m_{2\ell}.
\end{align}

\subsection{Pairwise dipole--dipole interaction}

The interaction energy between dipoles \(\mathbf m_{1k}\) and \(\mathbf m_{2\ell}\) is
\begin{equation}
	U_{k\ell}
	=
	\frac{\mu_0}{4\pi r_{k\ell}^3}
	\left[
	\mathbf m_{1k}\cdot \mathbf m_{2\ell}
	-
	3(\mathbf m_{1k}\cdot \hat{\mathbf r}_{k\ell})
	(\mathbf m_{2\ell}\cdot \hat{\mathbf r}_{k\ell})
	\right],
\end{equation}
where
\begin{equation}
	\mathbf r_{k\ell} = \mathbf r_{2\ell} - \mathbf r_{1k},
	\qquad
	r_{k\ell}=|\mathbf r_{k\ell}|,
	\qquad
	\hat{\mathbf r}_{k\ell}=\frac{\mathbf r_{k\ell}}{r_{k\ell}}.
\end{equation}

Writing
\begin{equation}
	\mathbf m_{1k}=m_{1k}\hat{\mathbf m}_{1k},
	\qquad
	\mathbf m_{2\ell}=m_{2\ell}\hat{\mathbf m}_{2\ell},
\end{equation}
we may express each term as
\begin{equation}
	U_{k\ell}
	=
	\frac{\mu_0}{4\pi}
	\left[
	\frac{A_{k\ell}\,m_{1k}m_{2\ell}}{r_{k\ell}^3}
	-
	\frac{3B_{k\ell}\,m_{1k}m_{2\ell}}{r_{k\ell}^5}
	\right],
\end{equation}
where
\begin{equation}
	A_{k\ell}=\hat{\mathbf m}_{1k}\cdot \hat{\mathbf m}_{2\ell},
	\qquad
	B_{k\ell}=(\hat{\mathbf m}_{1k}\cdot \mathbf r_{k\ell})(\hat{\mathbf m}_{2\ell}\cdot \mathbf r_{k\ell}).
\end{equation}

The total interaction energy is
\begin{equation}
	U_{\mathrm{int}}=\sum_{k=1}^{4}\sum_{\ell=1}^{4} U_{k\ell}.
\end{equation}

\subsection{All 16 pairwise contributions} \label{sec:4+4nonDeformedIndexCombos}

We now list the sixteen pair vectors \(\mathbf r_{k\ell}\), the scalar products \(A_{k\ell}\), and the quantities \(B_{k\ell}\).

\subsubsection*{\((k,\ell)=(1,1)\)}
\begin{align}
	\mathbf r_{11} &= (d-R_1+R_2 c)\hat y - R_2 s\,\hat z,\\
	r_{11}^2 &= (d-R_1+R_2 c)^2 + R_2^2 s^2,\\
	A_{11} &= 1,\\
	B_{11} &= 0,
\end{align}
hence
\begin{equation}
	U_{11}
	=
	\frac{\mu_0}{4\pi}\frac{m_{11}m_{21}}{r_{11}^3}.
\end{equation}

\subsubsection*{\((k,\ell)=(1,2)\)}
\begin{align}
	\mathbf r_{12} &= -R_2\hat x + (d-R_1)\hat y,\\
	r_{12}^2 &= R_2^2 + (d-R_1)^2,\\
	A_{12} &= 0,\\
	B_{12} &= R_2(R_1-d)c,
\end{align}
thus
\begin{equation}
	U_{12}
	=
	-\frac{3\mu_0}{4\pi}\,
	\frac{R_2(R_1-d)c\,m_{11}m_{22}}{r_{12}^5}.
\end{equation}

\subsubsection*{\((k,\ell)=(1,3)\)}
\begin{align}
	\mathbf r_{13} &= (d-R_1-R_2 c)\hat y + R_2 s\,\hat z,\\
	r_{13}^2 &= (d-R_1-R_2 c)^2 + R_2^2 s^2,\\
	A_{13} &= -1,\\
	B_{13} &= 0,
\end{align}
thus
\begin{equation}
	U_{13}
	=
	-\frac{\mu_0}{4\pi}\frac{m_{11}m_{23}}{r_{13}^3}.
\end{equation}

\subsubsection*{\((k,\ell)=(1,4)\)}
\begin{align}
	\mathbf r_{14} &= R_2\hat x + (d-R_1)\hat y,\\
	r_{14}^2 &= R_2^2 + (d-R_1)^2,\\
	A_{14} &= 0,\\
	B_{14} &= R_2(R_1-d)c,
\end{align}
hence
\begin{equation}
	U_{14}
	=
	-\frac{3\mu_0}{4\pi}\,
	\frac{R_2(R_1-d)c\,m_{11}m_{24}}{r_{14}^5}.
\end{equation}

\subsubsection*{\((k,\ell)=(2,1)\)}
\begin{align}
	\mathbf r_{21} &= R_1\hat x + (d+R_2 c)\hat y - R_2 s\,\hat z,\\
	r_{21}^2 &= R_1^2 + (d+R_2 c)^2 + R_2^2 s^2,\\
	A_{21} &= 0,\\
	B_{21} &= R_1(d+R_2 c),
\end{align}
hence
\begin{equation}
	U_{21}
	=
	-\frac{3\mu_0}{4\pi}\,
	\frac{R_1(d+R_2 c)\,m_{12}m_{21}}{r_{21}^5}.
\end{equation}

\subsubsection*{\((k,\ell)=(2,2)\)}
\begin{align}
	\mathbf r_{22} &= (R_1-R_2)\hat x + d\hat y,\\
	r_{22}^2 &= (R_1-R_2)^2 + d^2,\\
	A_{22} &= c,\\
	B_{22} &= c\,d^2,
\end{align}
thus
\begin{equation}
	U_{22}
	=
	\frac{\mu_0}{4\pi}
	\left(
	\frac{c\,m_{12}m_{22}}{r_{22}^3}
	-
	\frac{3c\,d^2\,m_{12}m_{22}}{r_{22}^5}
	\right).
\end{equation}

\subsubsection*{\((k,\ell)=(2,3)\)}
\begin{align}
	\mathbf r_{23} &= R_1\hat x + (d-R_2 c)\hat y + R_2 s\,\hat z,\\
	r_{23}^2 &= R_1^2 + (d-R_2 c)^2 + R_2^2 s^2,\\
	A_{23} &= 0,\\
	B_{23} &= R_1(R_2 c-d),
\end{align}
thus
\begin{equation}
	U_{23}
	=
	-\frac{3\mu_0}{4\pi}\,
	\frac{R_1(R_2 c-d)\,m_{12}m_{23}}{r_{23}^5}.
\end{equation}

\subsubsection*{\((k,\ell)=(2,4)\)}
\begin{align}
	\mathbf r_{24} &= (R_1+R_2)\hat x + d\hat y,\\
	r_{24}^2 &= (R_1+R_2)^2 + d^2,\\
	A_{24} &= -c,\\
	B_{24} &= -c\,d^2,
\end{align}
hence
\begin{equation}
	U_{24}
	=
	\frac{\mu_0}{4\pi}
	\left(
	-\frac{c\,m_{12}m_{24}}{r_{24}^3}
	+
	\frac{3c\,d^2\,m_{12}m_{24}}{r_{24}^5}
	\right).
\end{equation}

\subsubsection*{\((k,\ell)=(3,1)\)}
\begin{align}
	\mathbf r_{31} &= (d+R_1+R_2 c)\hat y - R_2 s\,\hat z,\\
	r_{31}^2 &= (d+R_1+R_2 c)^2 + R_2^2 s^2,\\
	A_{31} &= -1,\\
	B_{31} &= 0,
\end{align}
thus
\begin{equation}
	U_{31}
	=
	-\frac{\mu_0}{4\pi}\frac{m_{13}m_{21}}{r_{31}^3}.
\end{equation}

\subsubsection*{\((k,\ell)=(3,2)\)}
\begin{align}
	\mathbf r_{32} &= -R_2\hat x + (d+R_1)\hat y,\\
	r_{32}^2 &= R_2^2 + (d+R_1)^2,\\
	A_{32} &= 0,\\
	B_{32} &= R_2(d+R_1)c,
\end{align}
hence
\begin{equation}
	U_{32}
	=
	-\frac{3\mu_0}{4\pi}\,
	\frac{R_2(d+R_1)c\,m_{13}m_{22}}{r_{32}^5}.
\end{equation}

\subsubsection*{\((k,\ell)=(3,3)\)}
\begin{align}
	\mathbf r_{33} &= (d+R_1-R_2 c)\hat y + R_2 s\,\hat z,\\
	r_{33}^2 &= (d+R_1-R_2 c)^2 + R_2^2 s^2,\\
	A_{33} &= 1,\\
	B_{33} &= 0,
\end{align}
thus
\begin{equation}
	U_{33}
	=
	\frac{\mu_0}{4\pi}\frac{m_{13}m_{23}}{r_{33}^3}.
\end{equation}

\subsubsection*{\((k,\ell)=(3,4)\)}
\begin{align}
	\mathbf r_{34} &= R_2\hat x + (d+R_1)\hat y,\\
	r_{34}^2 &= R_2^2 + (d+R_1)^2,\\
	A_{34} &= 0,\\
	B_{34} &= R_2(d+R_1)c,
\end{align}
hence
\begin{equation}
	U_{34}
	=
	-\frac{3\mu_0}{4\pi}\,
	\frac{R_2(d+R_1)c\,m_{13}m_{24}}{r_{34}^5}.
\end{equation}

\subsubsection*{\((k,\ell)=(4,1)\)}
\begin{align}
	\mathbf r_{41} &= -R_1\hat x + (d+R_2 c)\hat y - R_2 s\,\hat z,\\
	r_{41}^2 &= R_1^2 + (d+R_2 c)^2 + R_2^2 s^2,\\
	A_{41} &= 0,\\
	B_{41} &= R_1(d+R_2 c),
\end{align}
thus
\begin{equation}
	U_{41}
	=
	-\frac{3\mu_0}{4\pi}\,
	\frac{R_1(d+R_2 c)\,m_{14}m_{21}}{r_{41}^5}.
\end{equation}

\subsubsection*{\((k,\ell)=(4,2)\)}
\begin{align}
	\mathbf r_{42} &= -(R_1+R_2)\hat x + d\hat y,\\
	r_{42}^2 &= (R_1+R_2)^2 + d^2,\\
	A_{42} &= -c,\\
	B_{42} &= -c\,d^2,
\end{align}
hence
\begin{equation}
	U_{42}
	=
	\frac{\mu_0}{4\pi}
	\left(
	-\frac{c\,m_{14}m_{22}}{r_{42}^3}
	+
	\frac{3c\,d^2\,m_{14}m_{22}}{r_{42}^5}
	\right).
\end{equation}

\subsubsection*{\((k,\ell)=(4,3)\)}
\begin{align}
	\mathbf r_{43} &= -R_1\hat x + (d-R_2 c)\hat y + R_2 s\,\hat z,\\
	r_{43}^2 &= R_1^2 + (d-R_2 c)^2 + R_2^2 s^2,\\
	A_{43} &= 0,\\
	B_{43} &= R_1(R_2 c-d),
\end{align}
thus
\begin{equation}
	U_{43}
	=
	-\frac{3\mu_0}{4\pi}\,
	\frac{R_1(R_2 c-d)\,m_{14}m_{23}}{r_{43}^5}.
\end{equation}

\subsubsection*{\((k,\ell)=(4,4)\)}
\begin{align}
	\mathbf r_{44} &= (R_2-R_1)\hat x + d\hat y,\\
	r_{44}^2 &= (R_2-R_1)^2 + d^2,\\
	A_{44} &= c,\\
	B_{44} &= c\,d^2,
\end{align}
hence
\begin{equation}
	U_{44}
	=
	\frac{\mu_0}{4\pi}
	\left(
	\frac{c\,m_{14}m_{24}}{r_{44}^3}
	-
	\frac{3c\,d^2\,m_{14}m_{24}}{r_{44}^5}
	\right).
\end{equation}

\subsection{Total interaction energy}

The total interaction energy between the two discrete toroidal systems is obtained by summing over all pairwise dipole interactions between the four loops of each torus. Using the dipole-dipole interaction formula introduced in Eq.~(S38), the total energy can be written as

\begin{equation}
	U_{\mathrm{int}}
	=
	\sum_{k=1}^{4}\sum_{\ell=1}^{4} U_{k\ell},
	\label{eq:Uint_sum}
\end{equation}

where each pair contribution \(U_{k\ell}\) is given by
\begin{equation}
	U_{k\ell}
	=
	\frac{\mu_0}{4\pi r_{k\ell}^3}
	\left[
	\mathbf m_{1k} \cdot \mathbf m_{2\ell}
	-
	3(\mathbf m_{1k} \cdot \hat{\mathbf r}_{k\ell})
	(\mathbf m_{2\ell} \cdot \hat{\mathbf r}_{k\ell})
	\right],
\end{equation}
with \(\mathbf r_{k\ell} = \mathbf r_{2\ell} - \mathbf r_{1k}\), \(r_{k\ell} = |\mathbf r_{k\ell}|\), and \(\hat{\mathbf r}_{k\ell} = \mathbf r_{k\ell}/r_{k\ell}\).

Substituting the explicit expressions for \(\mathbf r_{k\ell}\), \(A_{k\ell}\), and \(B_{k\ell}\) derived in Section~S2.4, the total interaction energy is fully determined by the sixteen contributions
\[
U_{\mathrm{int}} = U_{11} + U_{12} + \cdots + U_{44}.
\]

Equivalently, using the decomposition introduced in Eq.~(S41), one may write
\begin{equation}
	U_{\mathrm{int}}
	=
	\frac{\mu_0}{4\pi}
	\sum_{k=1}^{4}\sum_{\ell=1}^{4}
	\left[
	\frac{A_{k\ell}\, m_{1k} m_{2\ell}}{r_{k\ell}^3}
	-
	\frac{3 B_{k\ell}\, m_{1k} m_{2\ell}}{r_{k\ell}^5}
	\right],
	\label{eq:Uint_AB}
\end{equation}
where the coefficients \(A_{k\ell}\) and \(B_{k\ell}\) are explicitly listed for all sixteen pairs in the previous Section.

For identical loop magnitudes within each torus, one may further simplify by setting
\begin{equation}
	m_{11} = m_{12} = m_{13} = m_{14} \equiv m_1,
	\qquad
	m_{21} = m_{22} = m_{23} = m_{24} \equiv m_2,
\end{equation}
after which the structure of the interaction becomes more transparent and symmetry relations between different terms can be exploited.

\subsection{Remarks}
\begin{itemize}
	\item The interaction exhibits a nontrivial dependence on the relative angle $\alpha$ due to the rotation of the second torus.
	\item The 16 contributions fall into symmetry-related groups, which can be used to simplify the expression.
	\item In the limit $d \gg R_{1,2}$, the interaction reduces to an effective multipole interaction dominated by higher-order moments.
\end{itemize}
\subsection{Symmetry grouping of the 16 pair contributions} \label{sec:nonDeformedSymmetry}

For identical magnetic moment magnitudes within each torus,
\begin{equation}
	m_{11}=m_{12}=m_{13}=m_{14}\equiv m_1,
	\qquad
	m_{21}=m_{22}=m_{23}=m_{24}\equiv m_2,
\end{equation}
the sixteen pairwise dipole--dipole interaction terms can be grouped into symmetry-related classes.

We define the following distances:

\begin{align}
	D_+^2 &= (d - R_1 + R_2 \cos\alpha)^2 + R_2^2 \sin^2\alpha, \\
	D_-^2 &= (d - R_1 - R_2 \cos\alpha)^2 + R_2^2 \sin^2\alpha, \\
	E_+^2 &= (d + R_1 + R_2 \cos\alpha)^2 + R_2^2 \sin^2\alpha, \\
	E_-^2 &= (d + R_1 - R_2 \cos\alpha)^2 + R_2^2 \sin^2\alpha,
\end{align}

\begin{align}
	F_+^2 &= R_2^2 + (d + R_1)^2, \\
	F_-^2 &= R_2^2 + (d - R_1)^2,
\end{align}

\begin{align}
	G_+^2 &= R_1^2 + (d + R_2 \cos\alpha)^2 + R_2^2 \sin^2\alpha, \\
	G_-^2 &= R_1^2 + (d - R_2 \cos\alpha)^2 + R_2^2 \sin^2\alpha,
\end{align}

\begin{align}
	H_-^2 &= (R_1 - R_2)^2 + d^2, \\
	H_+^2 &= (R_1 + R_2)^2 + d^2.
\end{align}

The sixteen interaction terms organize into five symmetry classes:

\paragraph{Class A: axial pairs}
\begin{equation}
	U_A =
	\frac{\mu_0 m_1 m_2}{4\pi}
	\left(
	\frac{1}{D_+^{3}}
	-
	\frac{1}{D_-^{3}}
	-
	\frac{1}{E_+^{3}}
	+
	\frac{1}{E_-^{3}}
	\right).
\end{equation}

\paragraph{Class B: side pairs proportional to \(R_2 \cos\alpha\)}
\begin{equation}
	U_B =
	-\frac{3\mu_0 m_1 m_2}{4\pi}
	\left[
	\frac{2 R_2 (R_1 - d)\cos\alpha}{F_-^{5}}
	+
	\frac{2 R_2 (d + R_1)\cos\alpha}{F_+^{5}}
	\right].
\end{equation}

\paragraph{Class C: side pairs proportional to \(R_1\)}
\begin{equation}
	U_C =
	-\frac{3\mu_0 m_1 m_2}{4\pi}
	\left[
	\frac{2 R_1 (d + R_2 \cos\alpha)}{G_+^{5}}
	+
	\frac{2 R_1 (R_2 \cos\alpha - d)}{G_-^{5}}
	\right].
\end{equation}

\paragraph{Class D: coplanar diagonal pair with \((R_1 - R_2)\)}
\begin{equation}
	U_D =
	\frac{\mu_0 m_1 m_2}{4\pi}
	\left[
	\frac{2 \cos\alpha}{H_-^{3}}
	-
	\frac{6 d^2 \cos\alpha}{H_-^{5}}
	\right].
\end{equation}

\paragraph{Class E: coplanar diagonal pair with \((R_1 + R_2)\)}
\begin{equation}
	U_E =
	\frac{\mu_0 m_1 m_2}{4\pi}
	\left[
	-\frac{2 \cos\alpha}{H_+^{3}}
	+
	\frac{6 d^2 \cos\alpha}{H_+^{5}}
	\right].
\end{equation}

Therefore, the total interaction energy can be written as
\begin{equation}
	U_{\mathrm{int}} = U_A + U_B + U_C + U_D + U_E.
\end{equation}

This symmetry-grouped form is equivalent to the full sixteen-term sum, while making the dependence on the rotation angle \(\alpha\) explicit.
\begin{figure}[H] 
	\centering
	\includegraphics[width=0.85\textwidth]{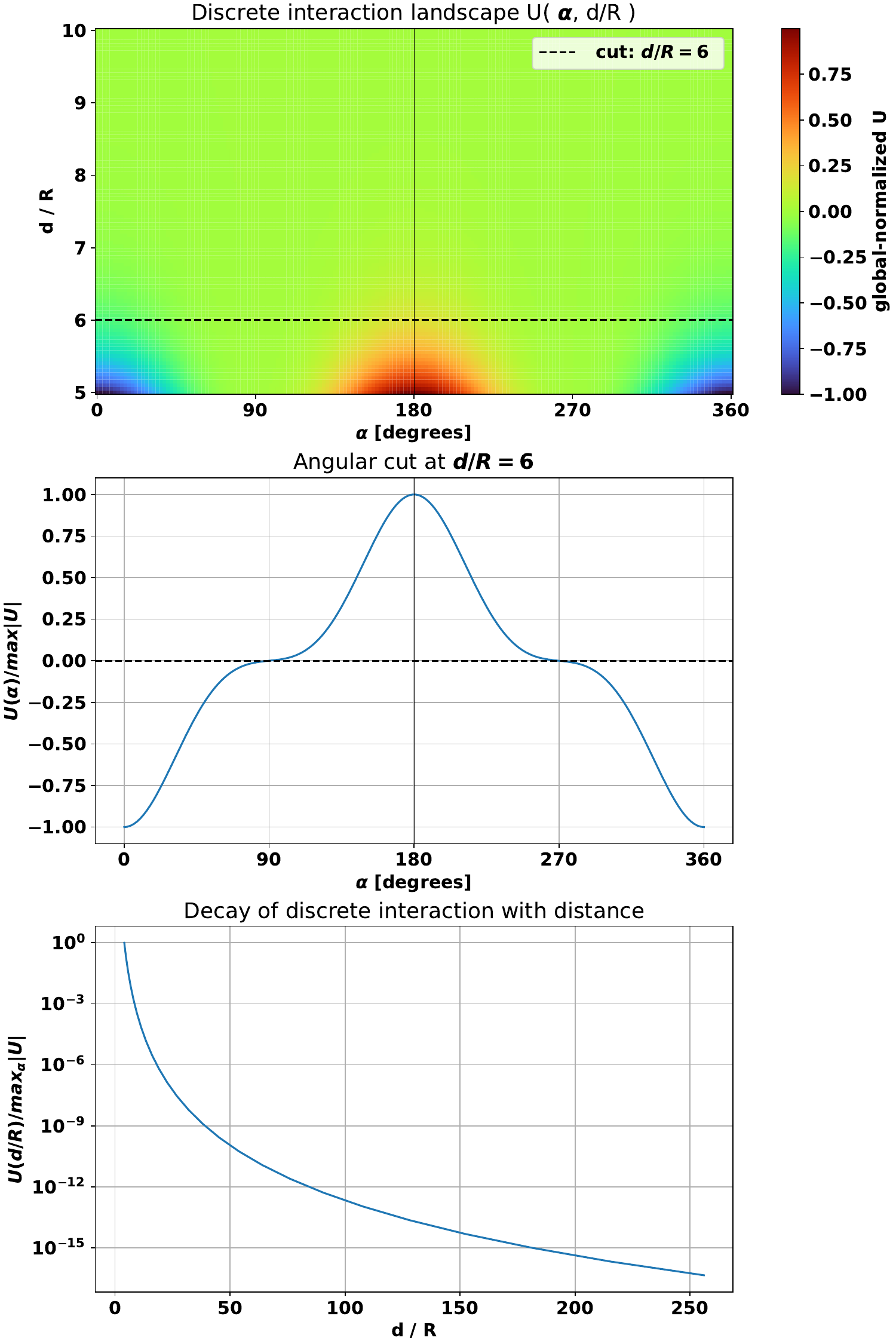}
	\caption{discrete interaction landscape $U(\alpha,d/R)$ for two non-deformed tori. 
		\textbf{Top:} Two-dimensional map of the interaction energy as a function of relative angle $\alpha$ 
		( for $\mathbf{ \hat{ y } }$ - rotation )
		and normalized separation $d/R$, globally normalized to $[-1,1]$. \\
		\textbf{Middle:} Angular dependence at fixed separation $d/R=6$. \\
		\textbf{Bottom:} Decay of the interaction strength $\max_{\alpha}|U|$ with distance, plotted on a logarithmic scale.}
	\label{fig:discreteNondefLandscapeY}
\end{figure}

Figure \ref{fig:discreteNondefLandscapeY} shows the interaction energy between the two discretely modeled tori as a function of their relative orientation and separation.

The energy landscape is smooth and depends symmetrically on the angle $\alpha$, showing a significant energy minimum ( thus strongest attractive force within the toroidal pair ) around $\alpha \approx 0$
( and $\alpha \approx 360^\circ$ ) where the toroidal moments are parallel;
the strongest repulsive force is situated around $\alpha \approx 180^\circ$ where they are antiparallel.

This landscape symmetry visualizes a geometric invariance of the model under
$\alpha \rightarrow 360^\circ - \alpha$ ( reverting the direction, but not severity of the relative tilt between the tori ).

The bottom panel presents a rapid decay of the interaction energy magnitude at increasing distance between the centers of the two tori.

\subsection{Elliptical torus}

We now consider a discrete model of two toroidal structures whose major rings are deformed into elliptical shapes. Each torus is represented by four magnetic dipoles placed at the principal points of the ellipse.
\begin{figure}[H]
	\centering
	\includegraphics[width=0.95\linewidth]{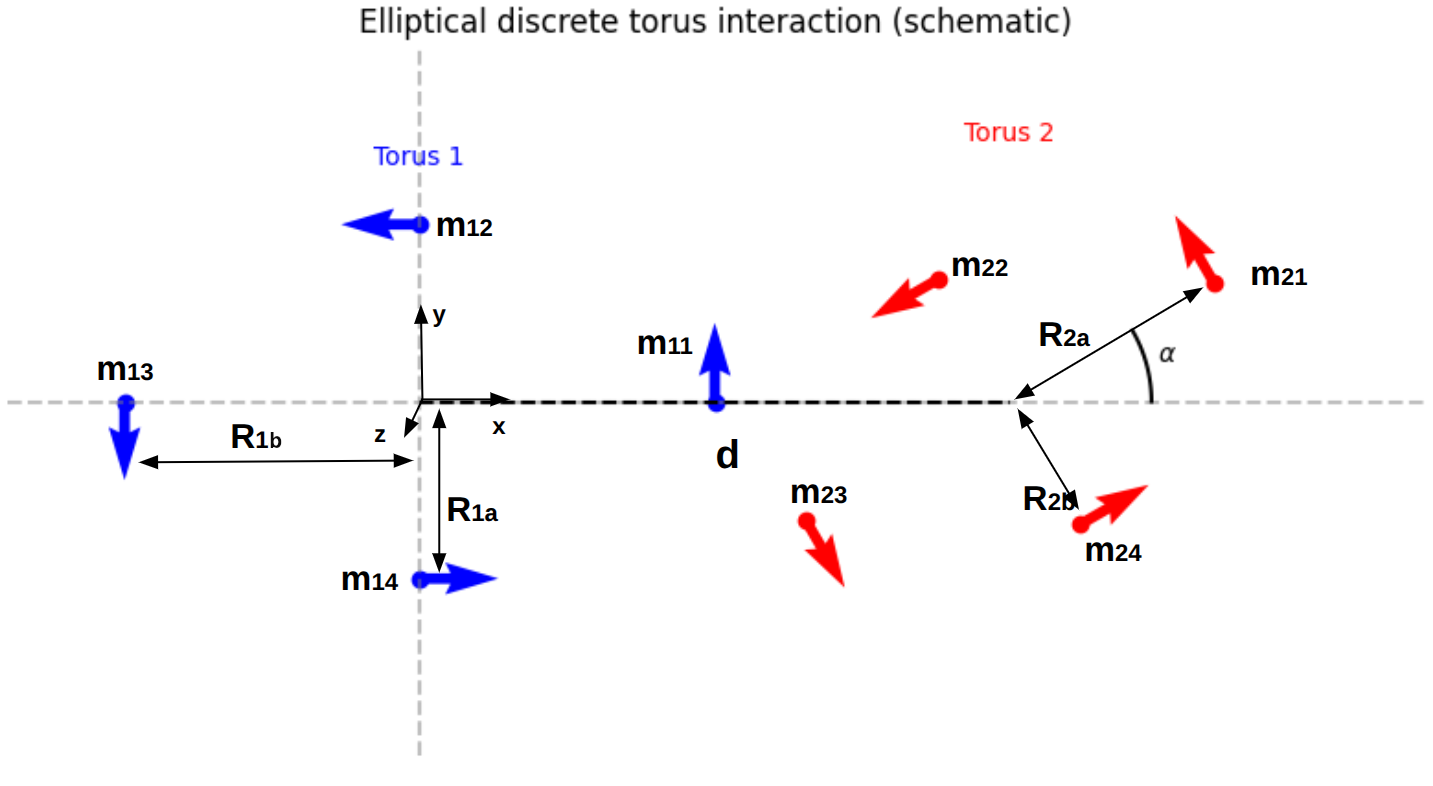}
	\caption{
		2 elliptical Torus Model
	}
	\label{fig:elltor}
\end{figure}
\subsubsection{Geometry of torus 1} \label{sec:discreteEllipticSpecs1}

The first torus is centered at the origin, with its four dipoles located at
\begin{align*}
	\mathbf r_{m_{11}} = (R_{1a}, 0),   & \quad 
	\mathbf r_{m_{12}} = (0, R_{1b}),   \\
	\mathbf r_{m_{13}} = (-R_{1a}, 0),  & \quad 
	\mathbf r_{m_{14}} = (0, -R_{1b}).
\end{align*}

The corresponding dipole orientations are taken to be tangential to the elliptical path:
\begin{equation}
	\hat{\mathbf m}_{11} = \hat{\mathbf y}, \quad 
	\hat{\mathbf m}_{12} = -\hat{\mathbf x}, \quad 
	\hat{\mathbf m}_{13} = -\hat{\mathbf y}, \quad 
	\hat{\mathbf m}_{14} = \hat{\mathbf x}.
\end{equation}

\subsubsection{Geometry of torus 2} \label{sec:discreteEllipticSpecs2}

The second torus is displaced by a distance $d$ along the $x$-axis and rotated by an angle $\alpha$ about the $z$-axis. Its dipole positions are given by
\begin{align}
	\mathbf r_{m_{21}} &= (d + R_{2a}\cos\alpha,\; R_{2a}\sin\alpha,\; 0), \\
	\mathbf r_{m_{22}} &= (d - R_{2b}\sin\alpha,\; R_{2b}\cos\alpha,\; 0), \\
	\mathbf r_{m_{23}} &= (d - R_{2a}\cos\alpha,\; -R_{2a}\sin\alpha,\; 0), \\
	\mathbf r_{m_{24}} &= (d + R_{2b}\sin\alpha,\; -R_{2b}\cos\alpha,\; 0).
\end{align}

The dipole orientations are obtained by rotating the corresponding tangential vectors:
\begin{align}
	\hat{\mathbf m}_{21} &= \sin\alpha\,\hat{\mathbf x} + \cos\alpha\,\hat{\mathbf y}, \\
	\hat{\mathbf m}_{22} &= -\cos\alpha\,\hat{\mathbf x} + \sin\alpha\,\hat{\mathbf y}, \\
	\hat{\mathbf m}_{23} &= -\sin\alpha\,\hat{\mathbf x} - \cos\alpha\,\hat{\mathbf y}, \\
	\hat{\mathbf m}_{24} &= \cos\alpha\,\hat{\mathbf x} - \sin\alpha\,\hat{\mathbf y}.
\end{align}

\subsubsection{Pairwise interaction}

The interaction energy between two dipoles $\mathbf m_{1k}$ and $\mathbf m_{2\ell}$ is given by the standard dipole--dipole interaction:
\begin{equation} \label{eq:standartMagneticInteraction}
	U_{k\ell}
	=
	\frac{\mu_0}{4\pi}
	\left[
	\frac{\hat{\mathbf m}_{1k} \cdot \hat{\mathbf m}_{2\ell}}{r_{k\ell}^3}
	-
	\frac{3(\hat{\mathbf m}_{1k} \cdot \mathbf r_{k\ell})(\hat{\mathbf m}_{2\ell} \cdot \mathbf r_{k\ell})}{r_{k\ell}^5}
	\right]
	m_{1k} m_{2\ell},
\end{equation}
where
\begin{equation}
	\mathbf r_{k\ell} = \mathbf r_{2\ell} - \mathbf r_{1k},
	\qquad
	r_{k\ell} = |\mathbf r_{k\ell}|.
\end{equation}

\subsubsection{Relative position vectors}

For each pair $(k,\ell)$, the relative vectors are explicitly computed. For example, for $k=1$:
\begin{align}
	\mathbf r_{11} &= (d + R_{2a}\cos\alpha - R_{1a})\,\hat{\mathbf x}
	+ (R_{2a}\sin\alpha)\,\hat{\mathbf y}, \\
	\mathbf r_{12} &= (d - R_{2b}\sin\alpha - R_{1a})\,\hat{\mathbf x}
	+ (R_{2b}\cos\alpha)\,\hat{\mathbf y}, \\
	\mathbf r_{13} &= (d - R_{2a}\cos\alpha - R_{1a})\,\hat{\mathbf x}
	- (R_{2a}\sin\alpha)\,\hat{\mathbf y}, \\
	\mathbf r_{14} &= (d + R_{2b}\sin\alpha - R_{1a})\,\hat{\mathbf x}
	- (R_{2b}\cos\alpha)\,\hat{\mathbf y}.
\end{align}

Analogous expressions are obtained for $k=2,3,4$.

\subsubsection{Total interaction energy}

The total interaction energy between the two elliptically deformed discrete tori is obtained by summing over all pairwise dipole--dipole interactions between the four dipoles of torus 1 and the four dipoles of torus 2. Thus,
\begin{equation}
	U_{\mathrm{tot}}
	=
	\sum_{k=1}^{4}\sum_{\ell=1}^{4} U_{k\ell},
\end{equation}
where each pair contribution is given by
\begin{equation}
	U_{k\ell}
	=
	\frac{\mu_0}{4\pi}
	\left[
	\frac{\mathbf m_{1k}\cdot \mathbf m_{2\ell}}{r_{k\ell}^{3}}
	-
	\frac{3(\mathbf m_{1k}\cdot \mathbf r_{k\ell})(\mathbf m_{2\ell}\cdot \mathbf r_{k\ell})}{r_{k\ell}^{5}}
	\right].
\end{equation}
Here
\begin{equation}
	\mathbf r_{k\ell}=\mathbf r_{2\ell}-\mathbf r_{1k},
	\qquad
	r_{k\ell}=|\mathbf r_{k\ell}|.
\end{equation}

Writing the dipole moments as
\begin{equation}
	\mathbf m_{1k}=m_{1k}\,\hat{\mathbf m}_{1k},
	\qquad
	\mathbf m_{2\ell}=m_{2\ell}\,\hat{\mathbf m}_{2\ell},
\end{equation}
the pair interaction may be rewritten as
\begin{equation}
	U_{k\ell}
	=
	\frac{\mu_0}{4\pi}
	\left[
	\frac{A_{k\ell}\,m_{1k}m_{2\ell}}{r_{k\ell}^{3}}
	-
	\frac{3B_{k\ell}\,m_{1k}m_{2\ell}}{r_{k\ell}^{5}}
	\right],
\end{equation}
where
\begin{equation}
	A_{k\ell}=\hat{\mathbf m}_{1k}\cdot \hat{\mathbf m}_{2\ell},
	\qquad
	B_{k\ell}=(\hat{\mathbf m}_{1k}\cdot \mathbf r_{k\ell})(\hat{\mathbf m}_{2\ell}\cdot \mathbf r_{k\ell}).
\end{equation}

Hence the total interaction energy takes the compact form
\begin{equation}
	U_{\mathrm{tot}}
	=
	\frac{\mu_0}{4\pi}
	\sum_{k=1}^{4}\sum_{\ell=1}^{4}
	\left[
	\frac{A_{k\ell}\,m_{1k}m_{2\ell}}{r_{k\ell}^{3}}
	-
	\frac{3B_{k\ell}\,m_{1k}m_{2\ell}}{r_{k\ell}^{5}}
	\right].
\end{equation}

For the elliptically deformed case, the rotational symmetry of the circular model is broken, and therefore the interaction can no longer be reduced to the five symmetry classes found in the undeformed geometry. One must instead evaluate all sixteen pair contributions explicitly.

In particular, the deformation enters through:
\begin{itemize}
	\item the modified dipole positions $\mathbf r_{1k}$ and $\mathbf r_{2\ell}$,
	\item the corresponding relative vectors $\mathbf r_{k\ell}$,
	\item the angle-dependent orientation factors $A_{k\ell}$,
	\item and the projection factors $B_{k\ell}$.
\end{itemize}

Thus, while the dipole--dipole interaction formula retains the same structure as in the circular case, the elliptical deformation makes the geometric dependence fully pair-specific.

\subsubsection{Remarks}

The elliptical deformation introduces anisotropy into the system through the distinct semi-axes $(R_{1a}, R_{1b})$ and $(R_{2a}, R_{2b})$. As a consequence:

\begin{itemize}
	\item The symmetry reduction from the circular case leads to a fully non-degenerate set of pairwise interactions.
	\item The angular dependence of the interaction becomes more intricate due to the coupling between $\alpha$ and the geometric anisotropy.
	\item The interaction energy depends sensitively on both the deformation parameters and the relative orientation.
\end{itemize}
\subsection{Discrete interaction of two symmetrically lopsided tori}

We consider now the interaction between two toroidal structures, both featuring a
smooth lopsided deformation of the inner radii
as described in main manuscript
and shown in fig. 2.

\subsubsection{Geometry of the Tori}

Consider the (\(4+4\)) case as defined in the main manuscript. The first torus is specified with

\begin{align}
	\hat{\mathbf m}_{11} &= -\hat x, & \mathbf r_{m_{11}} &= R_1 \hat y, &\\
	\hat{\mathbf m}_{12} &= -\hat y, & \mathbf r_{m_{12}} &= -R_1 \hat x, &\\
	\hat{\mathbf m}_{13} &= \hat x, & \mathbf r_{m_{13}} &= -R_1 \hat y, &\\
	\hat{\mathbf m}_{14} &= \hat y, & \mathbf r_{m_{14}} &= R_1 \hat x.
\end{align}

and the second one, centered at $d \hat{\mathbf{y}}$ and rotated in the $\hat{\mathbf{x}}$-direction  exactly as in section \ref{sec:4+4_nonDeformedSpecs}:

\begin{align}
	\hat{\mathbf m}_{21} &= -\hat x, & \mathbf r_{m_{21}} &= (d+R_2 c)\hat y - R_2 s\,\hat z,\\
	\hat{\mathbf m}_{22} &= -c\,\hat y + s\,\hat z, & \mathbf r_{m_{22}} &= -R_2 \hat x + d \hat y,\\
	\hat{\mathbf m}_{23} &= \hat x, & \mathbf r_{m_{23}} &= (d-R_2 c)\hat y + R_2 s\,\hat z,\\
	\hat{\mathbf m}_{24} &= c\,\hat y - s\,\hat z, & \mathbf r_{m_{24}} &= R_2 \hat x + d \hat y.
\end{align}

According to eq. ~6 in main manuscript, the magnetic moment of each current loop depends quadratically on its inner radius.
Therefore, we can model the deformation of the tori by varying the values of the magnetic moments representing the current loops.
We define six ( non-negative ) constants

\begin{equation}
	m_1, \quad m_{1a} \quad m_{1b}, \quad m_2, \quad m_{2a}, \quad m_{2b}.
\end{equation}

and set

\begin{align}
	m_{11} &= m_1,      & m_{21} &= m_2,  \\
	m_{12} &= m_{1a},   & m_{22} &= m_{2a},  \\
	m_{13} &= m_1,      & m_{23} &= m_2,  \\
	m_{14} &= m_{1b}.   & m_{24} &= m_{2b}.
\end{align}

By specifying the values of the six magnetic moments in the numerical approach later,
we match the deformation of the tori to the one described in section ~2.3 in main manuscript.

\subsubsection{Total interaction energy}

The calculation of the total interaction energy between the tori is entirely analogous to the non-deformed ($4+4$)-case and follows the same general formula

\begin{equation} \label{eq:4+4totalEnergy}
	U_{\mathrm{int}}
	=
	\sum_{k=1}^{4}\sum_{\ell=1}^{4}
	\frac{\mu_0}{4\pi r_{k\ell}^3}
	\left[
	\mathbf m_{1k} \cdot \mathbf m_{2\ell}
	-
	3(\mathbf m_{1k} \cdot \hat{\mathbf r}_{k\ell})
	(\mathbf m_{2\ell} \cdot \hat{\mathbf r}_{k\ell})
	\right]
\end{equation}

Again,
\begin{equation*}
	\mathbf{ r_{kl} } = \mathbf{ r_{ m_{ 2l } } }
	-
	\mathbf{ r_{ m_{ 1k } } }, r_{ kl } = | \mathbf{ r_{kl} } |.
\end{equation*}

Since we only vary the values of the magnetic moments but not their locations in this deformation case,
the exact values of $ r_{ k\ell } $ are the same as in section \ref{sec:4+4nonDeformedIndexCombos}.

\subsubsection{Symmetry grouping} \label{sec:smoothDefSymmetry}

As shown in section \ref{sec:nonDeformedSymmetry}, the 16 pairwise contributions to the total interaction energy between the two tori can be grouped according to the symmetries in the geometric arrangement of the individual magnetic moments.

If we split the 16 energy contributions into symmetry classes as described in section \ref{sec:nonDeformedSymmetry},
we receive

\begin{equation}
	U_A =
	\frac{\mu_0}{4\pi}
	\left(
	\frac{m_{1a} m_{2a}}{D_+^{3}}
	-
	\frac{m_{1a} m_{2b}}{D_-^{3}}
	-
	\frac{m_{1b} m_{2b}}{E_+^{3}}
	+
	\frac{m_{1b} m_{2a}}{E_-^{3}}
	\right).
\end{equation}

\begin{equation}
	U_B =
	-\frac{3\mu_0}{2\pi}
	\left[
	\frac{2 R_2 (R_1 - d)\cos\alpha (m_{1a} m_{2})}{F_-^{5}}
	+
	\frac{2 R_2 (d + R_1)\cos\alpha (m_{1b} m_{2})}{F_+^{5}}
	\right].
\end{equation}

\begin{equation}
	U_C =
	-\frac{3\mu_0}{2\pi}
	\left[
	\frac{R_1 (d + R_2 \cos\alpha) m_{1} m_{2a}}{G_+^{5}}
	+
	\frac{R_1 (R_2 \cos\alpha - d) m_{1} m_{2b}}{G_-^{5}}
	\right].
\end{equation}

\begin{equation}
	U_D =
	\frac{\mu_0 m_1 m_2}{2\pi}
	\left[
	\frac{\cos\alpha}{H_-^{3}}
	-
	\frac{3 d^2 \cos\alpha}{H_-^{5}}
	\right].
\end{equation}

\begin{equation}
	U_E =
	\frac{\mu_0 m_1 m_2}{2\pi}
	\left[
	-\frac{\cos\alpha}{H_+^{3}}
	+
	\frac{3 d^2 \cos\alpha}{H_+^{5}}
	\right].
\end{equation}

with

\begin{equation}
	U_{int} = U_A + U_B + U_C + U_D + U_E
\end{equation}

The classes largely resemble those of the non-deformed case in section \ref{sec:nonDeformedSymmetry}.
However, the loss of symmetry from the deformation of the tori ( varying $m_1$ and $m_2$ into three different variables each ) is apparent in $U_B$, $U_C$, and especially $U_A$, where a total of four different magnetic moments contribute to the energy calculation.

\subsection{Discrete interaction of two tori with alternating inner radii}

Now, we again evaluate the interaction between two deformed toroidal structures in the $(4+4)$-case,
but letting the inner radii of the current loops of both tori alternate between two constant values as illustrated in fig. 2.

\subsubsection{Geometry of the tori}

Once again we construct the (\(4+4\)) case as defined at the start of section \ref{sec:4+4_Introduction}. The first torus is specified with

\begin{align}
	\hat{\mathbf m}_{11} &= -\hat x, & \mathbf r_{m_{11}} &= R_1 \hat y, &\\
	\hat{\mathbf m}_{12} &= -\hat y, & \mathbf r_{m_{12}} &= -R_1 \hat x, &\\
	\hat{\mathbf m}_{13} &= \hat x, & \mathbf r_{m_{13}} &= -R_1 \hat y, &\\
	\hat{\mathbf m}_{14} &= \hat y, & \mathbf r_{m_{14}} &= R_1 \hat x.
\end{align}

and the second one, once more centered at $d \hat{\mathbf{y}}$ and rotated in the $\hat{\mathbf{x}}$-direction  exactly as in section \ref{sec:4+4_nonDeformedSpecs}:

\begin{align}
	\hat{\mathbf m}_{21} &= -\hat x, & \mathbf r_{m_{21}} &= (d+R_2 c)\hat y - R_2 s\,\hat z,\\
	\hat{\mathbf m}_{22} &= -c\,\hat y + s\,\hat z, & \mathbf r_{m_{22}} &= -R_2 \hat x + d \hat y,\\
	\hat{\mathbf m}_{23} &= \hat x, & \mathbf r_{m_{23}} &= (d-R_2 c)\hat y + R_2 s\,\hat z,\\
	\hat{\mathbf m}_{24} &= c\,\hat y - s\,\hat z, & \mathbf r_{m_{24}} &= R_2 \hat x + d \hat y.
\end{align}

We again model the deformation of the tori by varying the values of the magnetic moments representing each current loop.

For this we now define four magnetic moments

\begin{equation}
	m_{1c} \quad m_{1d}, \quad m_{2c}, \quad m_{2d}.
\end{equation}

and set

\begin{align}
	m_{11} &= m_{1c},   & m_{21} &= m_{2c},  \nonumber \\
	m_{12} &= m_{1d},   & m_{22} &= m_{2d},  \nonumber \\
	m_{13} &= m_{1c},   & m_{23} &= m_{2c},  \nonumber \\
	m_{14} &= m_{1d}.   & m_{24} &= m_{2d}.
	\label{eq:alternatingMomentAssignments}
\end{align}

\subsubsection{Total interaction energy}

The formula for calculating the total interaction energy between the two tori will again be equ. \ref{eq:4+4totalEnergy} with
\begin{equation*}
	\mathbf{ r_{kl} } = \mathbf{ r_{ m_{ 2l } } }
	-
	\mathbf{ r_{ m_{ 1k } } }, r_{ kl } = | \mathbf{ r_{kl} } |.
\end{equation*}

As the alternating loop size deformation model does not modify the locations of the magnetic moments, only their absolute values, the $ r_{ k\ell } $ are once more the exact same as in section \ref{sec:4+4nonDeformedIndexCombos}.

\subsubsection{Symmetry grouping}

In this deformation case we can once again recreate the symmetry grouping of the 16 pairwise energy contributions as defined in section \ref{sec:nonDeformedSymmetry}.

The energy contributions of the five groups are therefore

\begin{equation}
	U_A =
	\frac{\mu_0 m_{1c} m_{2c}}{4\pi}
	\left(
	\frac{1}{D_+^{3}}
	-
	\frac{1}{D_-^{3}}
	-
	\frac{1}{E_+^{3}}
	+
	\frac{1}{E_-^{3}}
	\right).
\end{equation}

\begin{equation}
	U_B =
	-\frac{3\mu_0 m_{1c} m_{2d}}{2\pi}
	\left[
	\frac{2 R_2 (R_1 - d)\cos\alpha}{F_-^{5}}
	+
	\frac{2 R_2 (d + R_1)\cos\alpha}{F_+^{5}}
	\right].
\end{equation}

\begin{equation}
	U_C =
	-\frac{3\mu_0 m_{1d} m_{2c}}{2\pi}
	\left[
	\frac{R_1 (d + R_2 \cos\alpha)}{G_+^{5}}
	+
	\frac{R_1 (R_2 \cos\alpha - d)}{G_-^{5}}
	\right].
\end{equation}

\begin{equation}
	U_D =
	\frac{\mu_0 m_{1d} m_{2d}}{2\pi}
	\left[
	\frac{\cos\alpha}{H_-^{3}}
	-
	\frac{3 d^2 \cos\alpha}{H_-^{5}}
	\right].
\end{equation}

\begin{equation}
	U_E =
	\frac{\mu_0 m_{1d} m_{2d}}{2\pi}
	\left[
	-\frac{\cos\alpha}{H_+^{3}}
	+
	\frac{3 d^2 \cos\alpha}{H_+^{5}}
	\right].
\end{equation}

\begin{equation}
	U_{int} = U_A + U_B + U_C + U_D + U_E
\end{equation}

Therefore the splitting of the two magnetic moments $m_1, m_2$ from section \ref{sec:nonDeformedSymmetry} into four individual moments again causes a loss of similarity between the formulas for $U_A$ to $U_E$ as it reduces the geometric symmetry of the two tori.
However, the loss of symmetry is not as drastic as in sec. \ref{sec:smoothDefSymmetry} since each of the five energies is calculated from only two different magnetic moments instead of three or four as was the case for $U_A$ to $U_C$ for the symmetrically lopsided deformation case.

\section{Tilt induced anisotropy}
\subsection{ Rectifying tilt } \label{sec:rectTilt}

We assume a non-deformed spin-toroidal structure of major radius $R$ made of $N$ magnetic moments of the same magnitude $m$.
Now, each moment $\mathbf{ m }_k, 1 \leq k \leq N$ is rotated by the same angle $\theta$ around the radial direction of the torus so that

\begin{align} \label{eq:rectMomentSplit}
	\hat{\mathbf {m}}_{k,rot}( \theta ) &= cos( \theta ) \hat{ \mathbf{ m } }_k
	+ sin( \theta ) \hat{ \mathbf{ z } }, &
\end{align}

The corresponding location vectors $\mathbf{r}_k = R \cdot \hat{\mathbf {r}}_k$ do not change as they represent the respective rotational axis of $\hat{\mathbf {m}}_{k,rot}$.

Therefore the toroidal moment of the structure is

\begin{align} \label{eq:rectToroidalMoment}
	\mathbf{ T }    &= \frac{1}{2} \sum_{ k=1 }^N \mathbf{r}_k \times \mathbf {m}_{k,rot}( \theta ) \nonumber \\
	&= \frac{Rm}{2} \sum_{ k=1 }^N \hat{\mathbf{r}}_k
	\times \hat{\mathbf {m}}_{k,rot}( \theta ) \nonumber \\
	&= \frac{Rm}{2} \sum_{ k=1 }^N \hat{\mathbf{r}}_k
	\times ( \cos( \theta ) \hat{ \mathbf{ m } }_k
	+ sin( \theta ) \hat{ \mathbf{ z } } ) \nonumber \\
	&= \frac{NRm \cos( \theta )}{2} \hat{ \mathbf{ z } } + \frac{Rm}{2} \sum_{ k=1 }^N \hat{\mathbf{r}}_k
	\times sin( \theta ) \hat{ \mathbf{ z } } \nonumber \\
	&= \cos( \theta ) \mathbf{ T }_0 + 0 \nonumber \\
	&= \cos( \theta ) \mathbf{ T }_0
\end{align}

In this calculation, $\mathbf{ T }_0$ represents the toroidal moment of an entirely non-deformed torus as introduced in Sec.~2.1 of the main manuscript.
The remaining sum evaluates to 0 since the $\hat{\mathbf{r}}_k$ are uniformly distributed directions within the $\hat{ \mathbf{ x } } \hat{ \mathbf{ y } }$-plane.

The residual ( far-field ) magnetic moment evaluates to

\begin{align} \label{eq:rectMagneticMoment}
	\mathbf{m}_{\mathrm{tot}}
	&=
	\sum_{k=1}^N \mathbf{m}_{k,rot} \nonumber \\
	&=
	m \sum_{k=1}^N cos( \theta ) \hat{ \mathbf{ m } }_k + sin( \theta ) \hat{ \mathbf{ z } } \nonumber \\
	&=
	Nm sin( \theta ) \hat{ \mathbf{ z } }
\end{align}

using equation \ref{eq:rectMomentSplit} and the $\hat{ \mathbf{ m } }_k$ again being evenly distributed directions within the $\hat{ \mathbf{ x } } \hat{ \mathbf{ y } }$-plane.

As a sanity check for our results, we set $\theta = 0$ and get $\mathbf{ T } = \mathbf{ T }_0$ and $\mathbf{m}_{\mathrm{tot}} = 0$ which are the correct results for a perfectly non-deformed torus.
As a contrast, one may set $\theta = 90^\circ$ and find 
\begin{equation*}
	\mathbf{ T } = 0, \quad \mathbf{m}_{\mathrm{tot}} = Nm \hat{ \mathbf{ z } }.
\end{equation*}

In this sense, increasing $\theta$ from $0^\circ$ to $90^\circ$ has a rectifying effect on the magnetic moments, transitioning their arrangement from a perfectly toroidal to a perfectly magnetic dipole, both their dipole moments parallel to one another ( $\hat{ \mathbf{ z }}$ ).

\subsubsection{Internal potential energy of a partially rectified torus}

To describe the energetic cost associated with deviations from the ideal
toroidal configuration, we introduce a simple phenomenological anisotropy
energy of the form
\[
E(\theta)=E_0+K\sin^2\theta,
\]
where $E_0$ denotes the energy of the ideal toroidal state and $K$ represents
an effective anisotropy energy scale. This form captures the gradual suppression
of toroidal order as the local magnetic moments tilt away from their ideal
tangential arrangement.

\subsection{Pairwise tilt}

Another scenario of interest is to tilt the $\mathbf{ m }_k$ from the tangential direction but in such a manner that the total residual magnetic moment is always zero as was discussed to be ideal in main manuscript.
Starting with the same non-deformed model as in section \ref{sec:rectTilt}, the magnetic moment locations are again unchanged while the magnetic moments themselves tilt in a pairwise manner
in the $\hat{ \mathbf{ x } } \hat{ \mathbf{ y } }$-plane:

\begin{align} \label{eq:pairTiltMomentSplit}
	\hat{\mathbf {m}}_{k,rot}( \alpha_k ) 
	&= 
	\cos( \alpha_k ) \hat{\mathbf {m}}_k + \sin( \alpha_k ) \hat{\mathbf {r}}_k, \nonumber \\
	\hat{\mathbf {m}}_{ k + \frac{N}{2},rot}( \alpha_k )
	&=
	\cos( \alpha_k ) \hat{\mathbf {m}}_{ k + \frac{N}{2} } 
	+ \sin( \alpha_k ) \hat{\mathbf {r}}_{ k + \frac{N}{2} }.
\end{align}

Since we set $N$ to be even, this pair of ( directly opposing ) initial magnetic moments and their locations will be exact negatives of one another. Therefore, the sum of the two altered magnetic moments

\begin{equation}
	\hat{\mathbf {m}}_{k,rot}( \alpha_k ) + \hat{\mathbf {m}}_{ k + \frac{N}{2},rot}( \alpha_k ) = 0
	\quad \forall k
\end{equation}

satisfies the premise that the residual magnetic moment should not change from ( pairwise ) tilting.
Because the perfectly regular, non-deformed toroidal arrangement of magnetic moments satisfies

\begin{equation} \label{eq:pairTiltMagneticMoment}
	\mathbf{m}_{\mathrm{tot}} \equiv 0
\end{equation}

according to our calculations in main text, equ. \ref{eq:pairTiltMagneticMoment} will be satisfied for any values of the $\alpha_k$.
Note that the set of $\alpha_k, \quad k = 1, ... \frac{N}{2}$ contains $\frac{N}{2}$ independent tilting angles.

The total toroidal moment evaluates to

\begin{align} \label{eq:pairTiltToroidalMoment}
	\mathbf{ T }    &= \frac{1}{2} \sum_{ k=1 }^N \mathbf{r}_k 
	\times \mathbf {m}_{k,rot}( \alpha_k ) \nonumber \\
	&= \frac{1}{2} \sum_{ k=1 }^{ \frac{N}{2} } \mathbf{r}_k 
	\times \mathbf {m}_{k,rot}( \alpha_k )
	+ \mathbf{r}_{ k + \frac{N}{2} }
	\times \mathbf {m}_{ { k + \frac{N}{2} } ,rot}( \alpha_k )\nonumber \\
	&= \frac{1}{2} \sum_{ k=1 }^{ \frac{N}{2} } \mathbf{r}_k 
	\times \mathbf {m}_{k,rot}( \alpha_k )
	+ ( - \mathbf{r}_k
	\times ( - \mathbf {m}_{k,rot}( \alpha_k ))) \nonumber \\
	&= \sum_{ k=1 }^{ \frac{N}{2} } \mathbf{r}_k 
	\times \mathbf {m}_{k,rot}( \alpha_k ) \nonumber \\
	&= Rm \sum_{ k=1 }^{ \frac{N}{2} } \hat{ \mathbf{r}_k }
	\times 
	( \mathbf \cos( \alpha_k ) \hat{\mathbf {m}}_k 
	+ \sin( \alpha_k ) \hat{\mathbf {r}_k}) \nonumber \\
	&= Rm \sum_{ k=1 }^{ \frac{N}{2} } 
	\mathbf \cos( \alpha_k ) \hat{ \mathbf{ z } }.
\end{align}

Note that the sum cannot be completely simplified as the $\alpha_k$ are generally set independently from one another.
In this scenario, the sanity check $\alpha_k = 0 \quad \forall k$ also gives

\begin{equation}
	\mathbf{ T } = \mathbf{ T }_0 = \frac{NRm}{2} \hat{ \mathbf{ z } }
	\quad \mathbf{m}_{\mathrm{tot}} = 0
\end{equation}

as expected for the perfectly non-deformed torus.


{\color{black}
	\section{Discretization, symmetry, and far-field multipole analysis}
	\label{sec:SI_convergence_multipoles}
	
	The interaction evaluated throughout the manuscript is the conventional pairwise magnetic dipole--dipole interaction between localized magnetic moments. The toroidal moment characterizes the collective vortex-like arrangement of these moments, but it does not by itself define an additional elementary interaction term. To distinguish features associated with the toroidal texture from those caused by the finite $4+4$ discretization, we generalize the ideal ring to $N$ equally spaced tangential magnetic moments.
	
	\subsection{Fixed-toroidal-moment normalization}
	
	For an ideal circular ring of radius $R$ with $N$ equal tangential moments of magnitude $m_N$, the toroidal moment is
	\begin{equation}
		T_z=\frac{1}{2}NRm_N.
	\end{equation}
	To compare different discretizations without changing the total toroidal moment, we therefore impose
	\begin{equation}
		m_N=\frac{2T_0}{NR},
		\label{eq:SI_fixedTnorm}
	\end{equation}
	so that $T_z=T_0$ for every value of $N$. The total magnetic dipole moment remains zero.
	
	\subsection{Fourier selection rule from $C_N$ symmetry}
	
	In general, the angular dependence of the interaction energy may be written as
	\begin{equation}
		U(\alpha)=U_0+\sum_{n=1}^{\infty}
		\left[
		A_n\cos(n\alpha)+B_n\sin(n\alpha)
		\right].
		\label{eq:fourier_interaction}
	\end{equation}
	The allowed Fourier components are constrained by the rotational symmetry
	of the discrete magnetic configuration.
	
	For the ideal $N$-site ring, the site angles are
	\begin{equation}
		\phi_k=\frac{2\pi k}{N},\qquad k=0,\ldots,N-1.
	\end{equation}
	Any azimuthal multipole component contains the roots-of-unity sum
	\begin{equation}
		S_m=\sum_{k=0}^{N-1}e^{-im\phi_k}
		=\sum_{k=0}^{N-1}e^{-i2\pi m k/N}.
	\end{equation}
	Using the finite geometric series,
	\begin{equation}
		S_m=
		\begin{cases}
			N, & m=qN,\quad q\in\mathbb{Z},\\
			0, & \text{otherwise}.
		\end{cases}
		\label{eq:SI_rootunity}
	\end{equation}
	Thus, the $C_N$ symmetry removes all azimuthal components except integer multiples of $N$. For tangential magnetic moments, the $m=0$ contribution vanishes because the angular derivative introduces an explicit factor of $m$. Hence the first allowed azimuthal components are
	\begin{equation}
		m=\pm N.
	\end{equation}
	Under a relative rotation of the second ring by an angle $\alpha$, these components acquire phases $e^{\mp iN\alpha}$. Their real combination therefore gives the leading angular dependence
	\begin{equation}
		U_N(\alpha)\propto \cos(N\alpha).
		\label{eq:SI_cosNalpha}
	\end{equation}
	This explains the $N$-fold angular periodicity observed numerically.
	
	\subsection{Leading multipole order and far-field scaling}
	
	The external magnetic scalar potential can be expanded in spherical harmonics as
	\begin{equation}
		\Phi(\mathbf r)
		=\sum_{\ell,m}\frac{Q_{\ell m}}{r^{\ell+1}}Y_{\ell m}(\theta,\phi),
	\end{equation}
	where the multipole coefficients for the tangential point-dipole ring inherit the selection rule in Eq.~\eqref{eq:SI_rootunity}. Since $|m|\leq \ell$ and the first allowed nonzero value is $|m|=N$, the first non-vanishing external multipole occurs at
	\begin{equation}
		\ell_{\min}=N.
		\label{eq:SI_lminN}
	\end{equation}
	The interaction between multipoles of orders $\ell_1$ and $\ell_2$ scales asymptotically as
	\begin{equation}
		U_{\ell_1\ell_2}(d)\propto d^{-(\ell_1+\ell_2+1)}.
	\end{equation}
	For two identical ideal $N$-site rings, $\ell_1=\ell_2=N$, and therefore
	\begin{equation}
		U_N(d)\propto d^{-(2N+1)}.
		\label{eq:SI_farfieldlaw}
	\end{equation}
	The absence of a $d^{-3}$ term is consistent with the vanishing net magnetic dipole moment. As $N$ increases, progressively more low-order multipoles are eliminated by symmetry, making the interaction extremely short-ranged even when the total toroidal moment is held fixed.
	
	\subsection{High-precision numerical verification}
	
	We evaluated the ideal-ring interaction with arbitrary-precision arithmetic for $N=4,8,16,$ and $32$, using the normalization in Eq.~\eqref{eq:SI_fixedTnorm}. The fitted far-field exponents are summarized in Table~\ref{tab:SI_farfield_exponents}.
	
	{\color{black}
		Figure~S3 illustrates the effect of refining the discrete ring while keeping
		the total toroidal moment fixed. The unnormalized interaction shows a rapid
		suppression of the interaction amplitude with increasing $N$. At the same
		time, the normalized angular profiles exhibit an increasing number of
		oscillations, reflecting the $C_N$ rotational symmetry of the discrete
		configuration. Thus, refinement of the discretization does not simply
		reproduce the $N=4$ landscape with greater resolution; rather, it
		systematically shifts the interaction to higher angular harmonics while
		strongly suppressing its magnitude.
	}
	
	\begin{table}[H]
		\centering
		\caption{High-precision far-field exponents for ideal discrete spin-toroidal rings. The fitted exponents agree with the analytic prediction $p_N=2N+1$.}
		\label{tab:SI_farfield_exponents}
		\begin{tabular}{c c c}
			\hline
			$N$ & fitted $p_N$ & $2N+1$ \\
			\hline
			4  & 9.0291  & 9  \\
			8  & 17.0359 & 17 \\
			16 & 33.0601 & 33 \\
			32 & 65.0808 & 65 \\
			\hline
		\end{tabular}
	\end{table}
	
	\begin{figure}[H]
		\centering
		\begin{minipage}{0.48\textwidth}
			\centering
			\includegraphics[width=\linewidth]{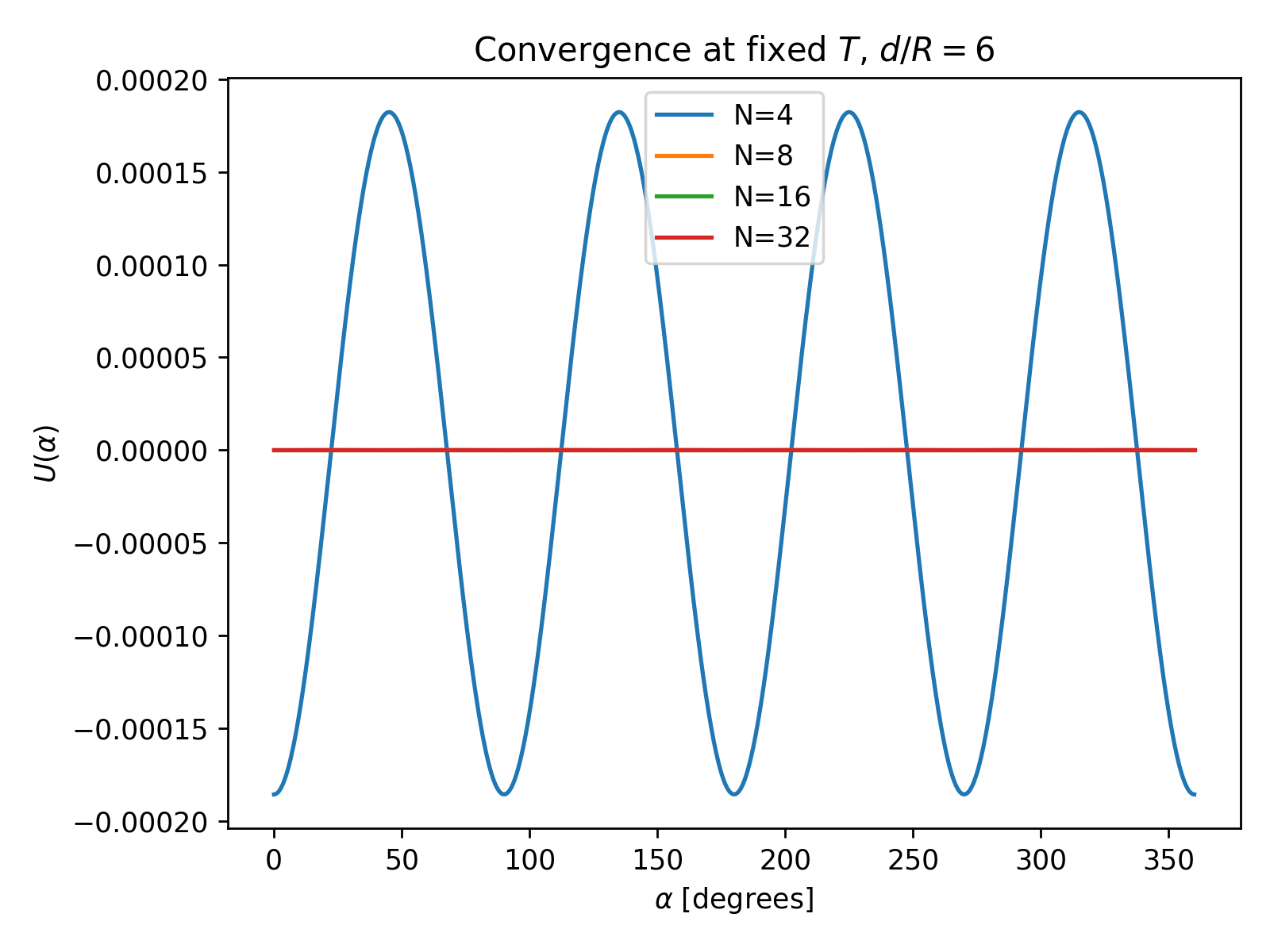}
		\end{minipage}\hfill
		\begin{minipage}{0.48\textwidth}
			\centering
			\includegraphics[width=\linewidth]{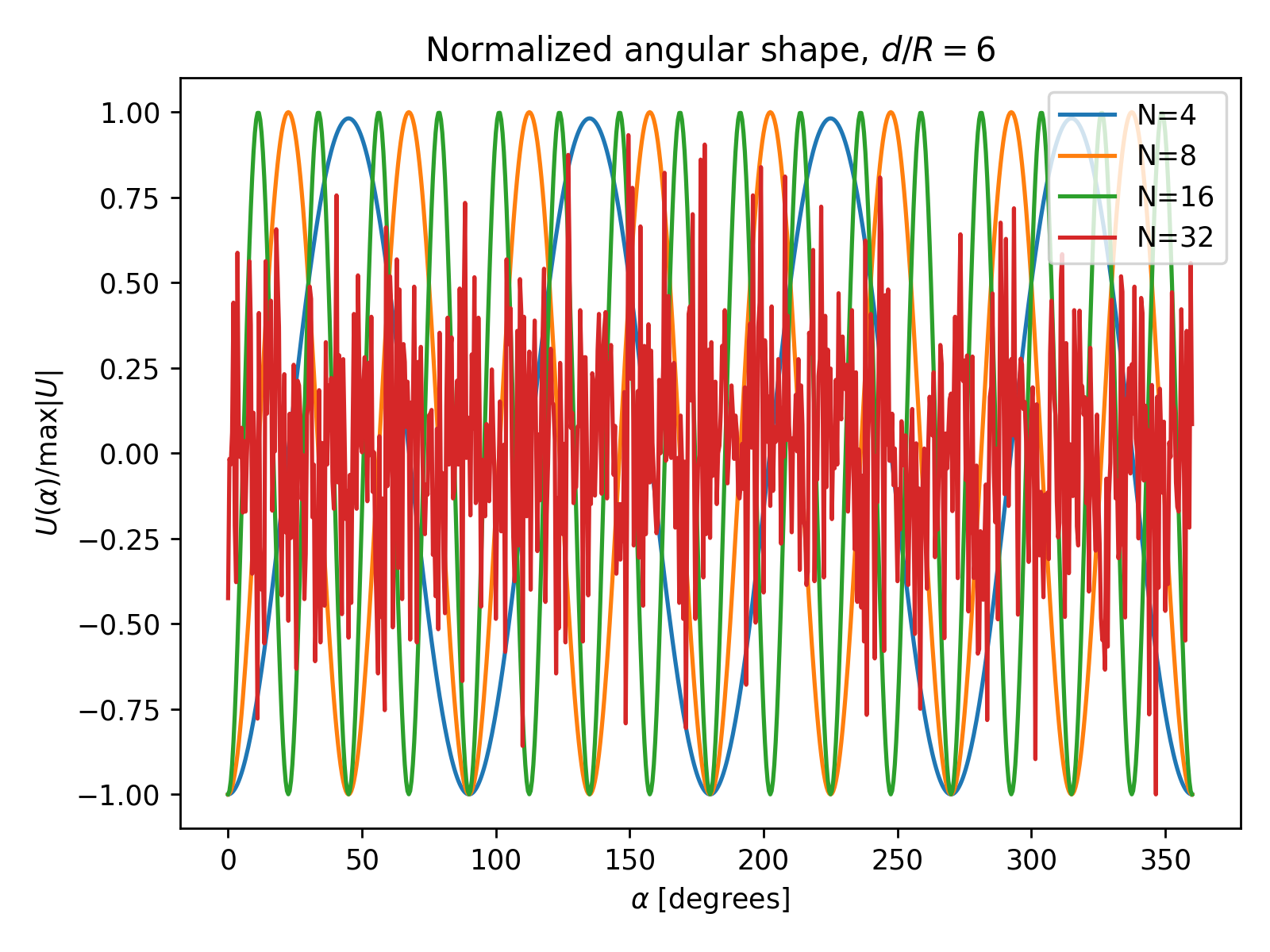}
		\end{minipage}
		\caption{Discretization study at fixed total toroidal moment. Left: raw angular interaction $U_N(\alpha)$ for $N=4,8,16,$ and $32$. Right: the corresponding normalized angular profiles. The rapidly decreasing absolute interaction amplitude with increasing $N$ demonstrates that the interaction is not determined by the toroidal moment alone.}
		\label{fig:SI_convergence}
	\end{figure}
	
	\begin{figure}[H]
		\centering
		\begin{minipage}{0.48\textwidth}
			\centering
			\includegraphics[width=\linewidth]{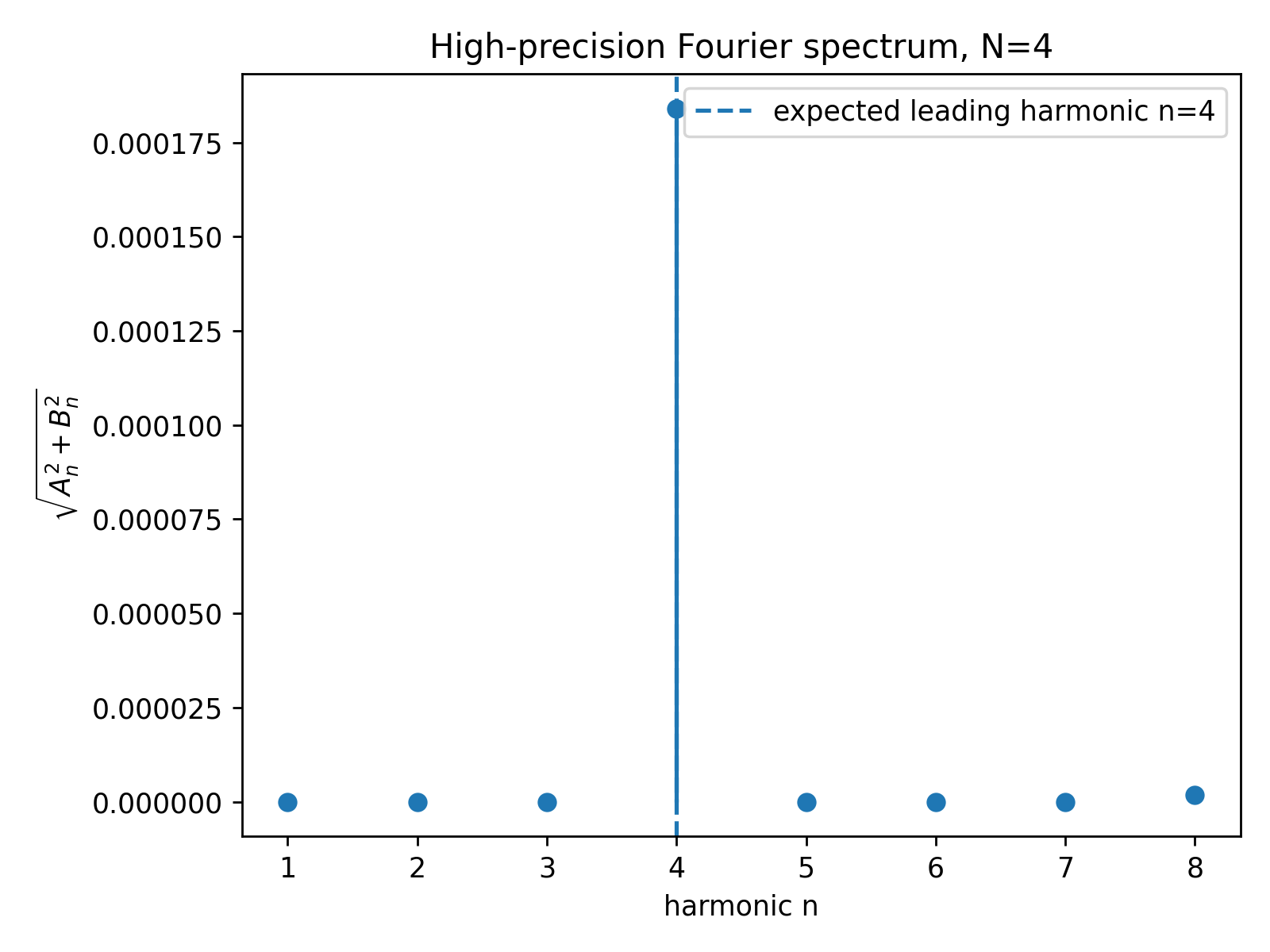}\\[-1mm]
			$N=4$
		\end{minipage}\hfill
		\begin{minipage}{0.48\textwidth}
			\centering
			\includegraphics[width=\linewidth]{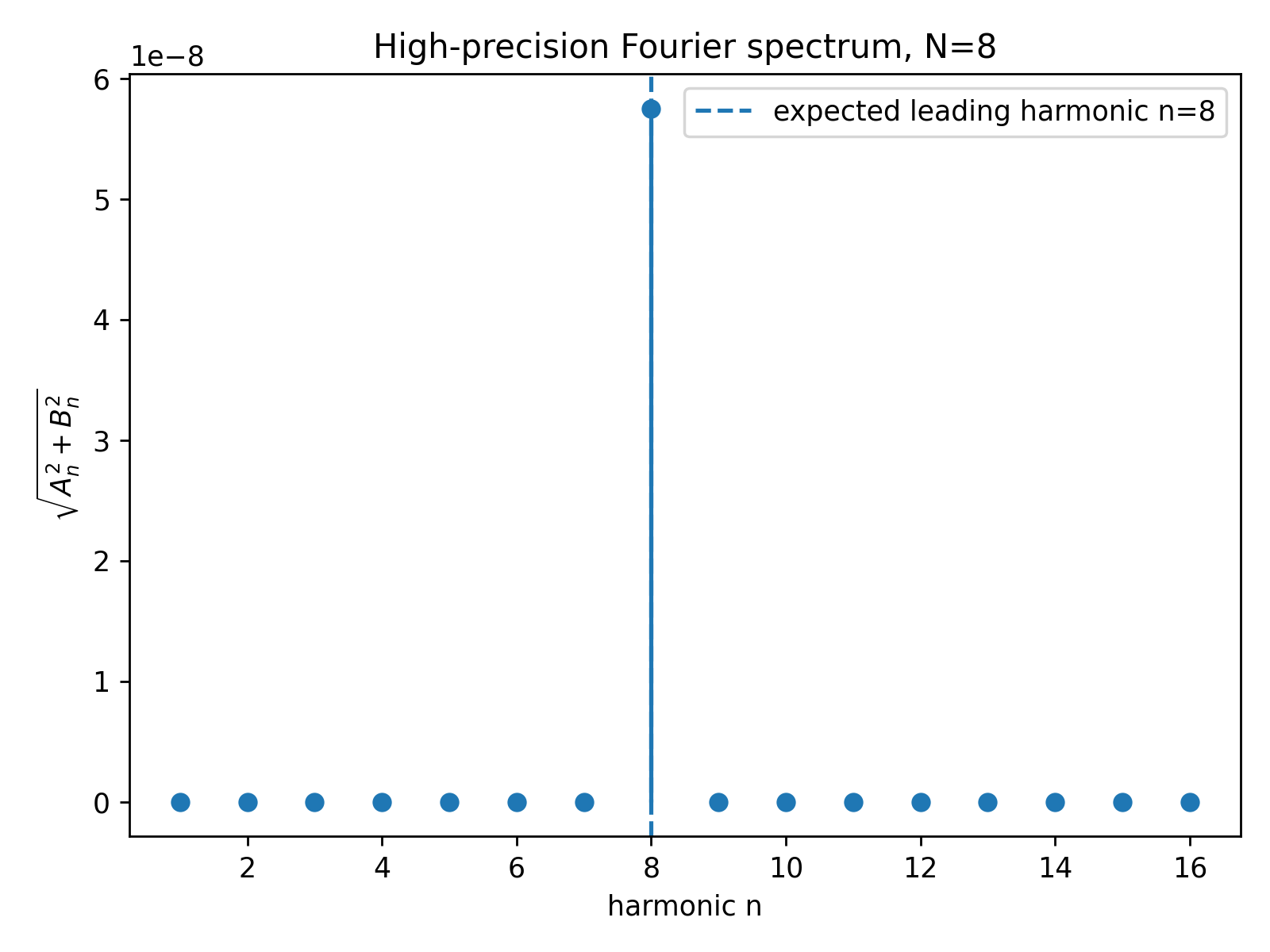}\\[-1mm]
			$N=8$
		\end{minipage}
		
		\vspace{2mm}
		
		\begin{minipage}{0.48\textwidth}
			\centering
			\includegraphics[width=\linewidth]{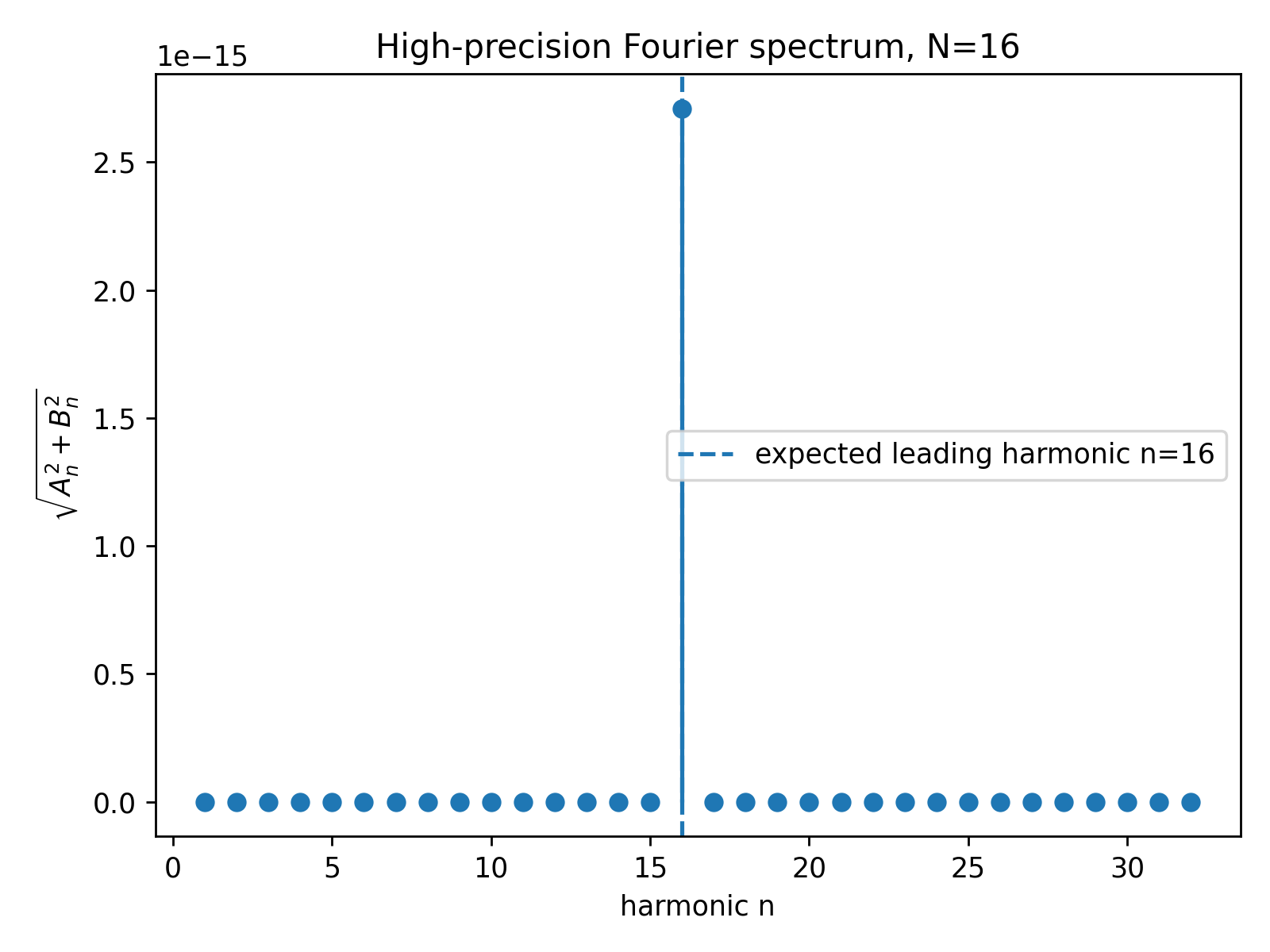}\\[-1mm]
			$N=16$
		\end{minipage}\hfill
		\begin{minipage}{0.48\textwidth}
			\centering
			\includegraphics[width=\linewidth]{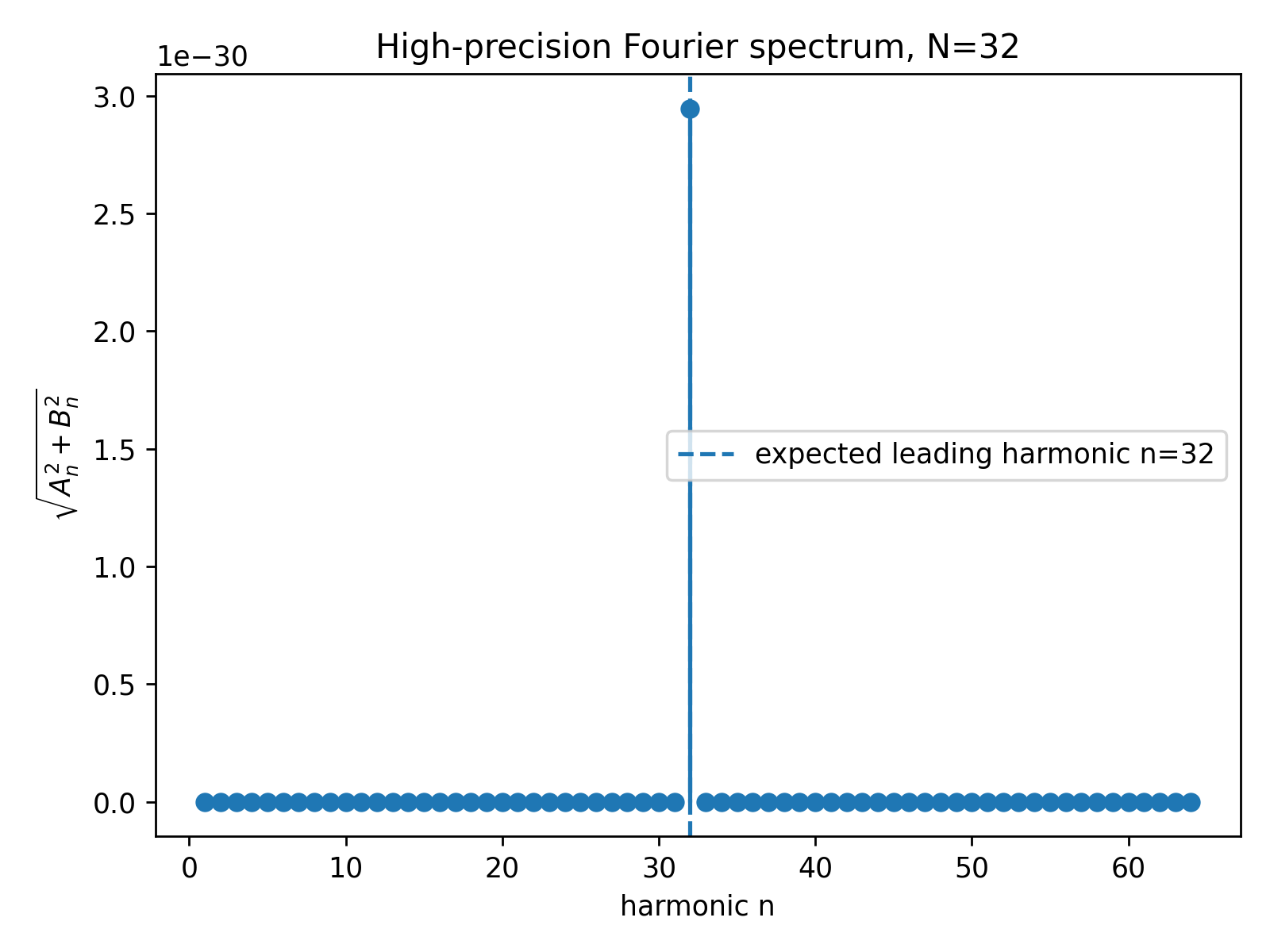}\\[-1mm]
			$N=32$
		\end{minipage}
		\caption{High-precision Fourier spectra of the ideal $N$-site interaction at fixed separation. In every case the leading harmonic occurs at $n=N$, in agreement with the $C_N$ selection rule.}
		\label{fig:SI_fourier_spectra}
	\end{figure}
	{\color{black}
		The Fourier spectra provide a direct numerical verification of the symmetry
		selection rule derived in Sec.~S4.2. For each discretization, the first and
		dominant Fourier component occurs at $n=N$, while the lower harmonics vanish.
		Thus,
		\[
		N=4,8,16,32
		\quad\Longrightarrow\quad
		n_{\mathrm{lead}}=4,8,16,32,
		\]
		respectively. This establishes that the characteristic angular periodicity of
		the interaction landscape originates from the discrete $C_N$ symmetry of the
		underlying magnetic-moment arrangement.
	}
	\begin{figure}[H]
		\centering
		\begin{minipage}{0.48\textwidth}
			\centering
			\includegraphics[width=\linewidth]{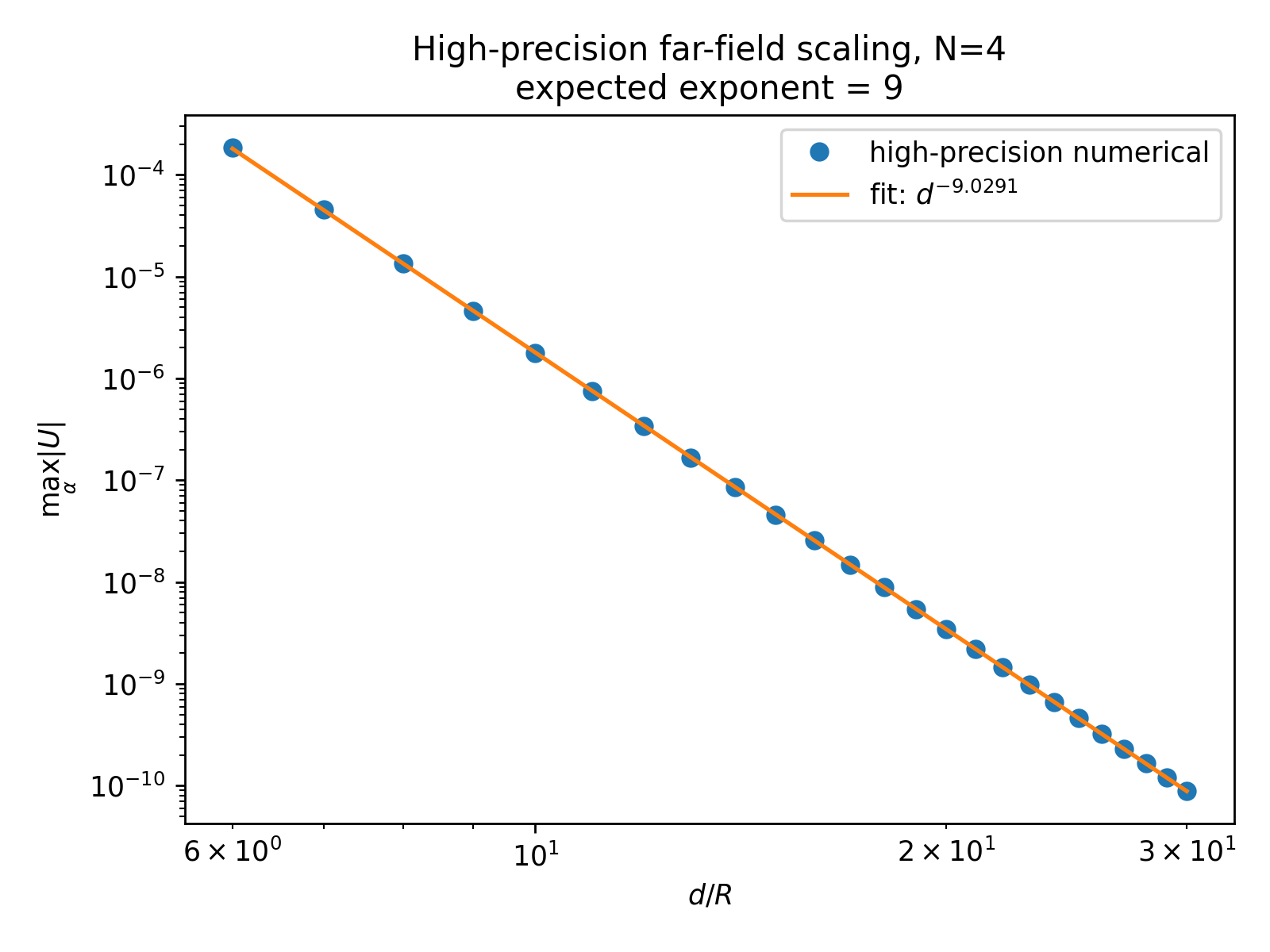}\\[-1mm]
			$N=4$
		\end{minipage}\hfill
		\begin{minipage}{0.48\textwidth}
			\centering
			\includegraphics[width=\linewidth]{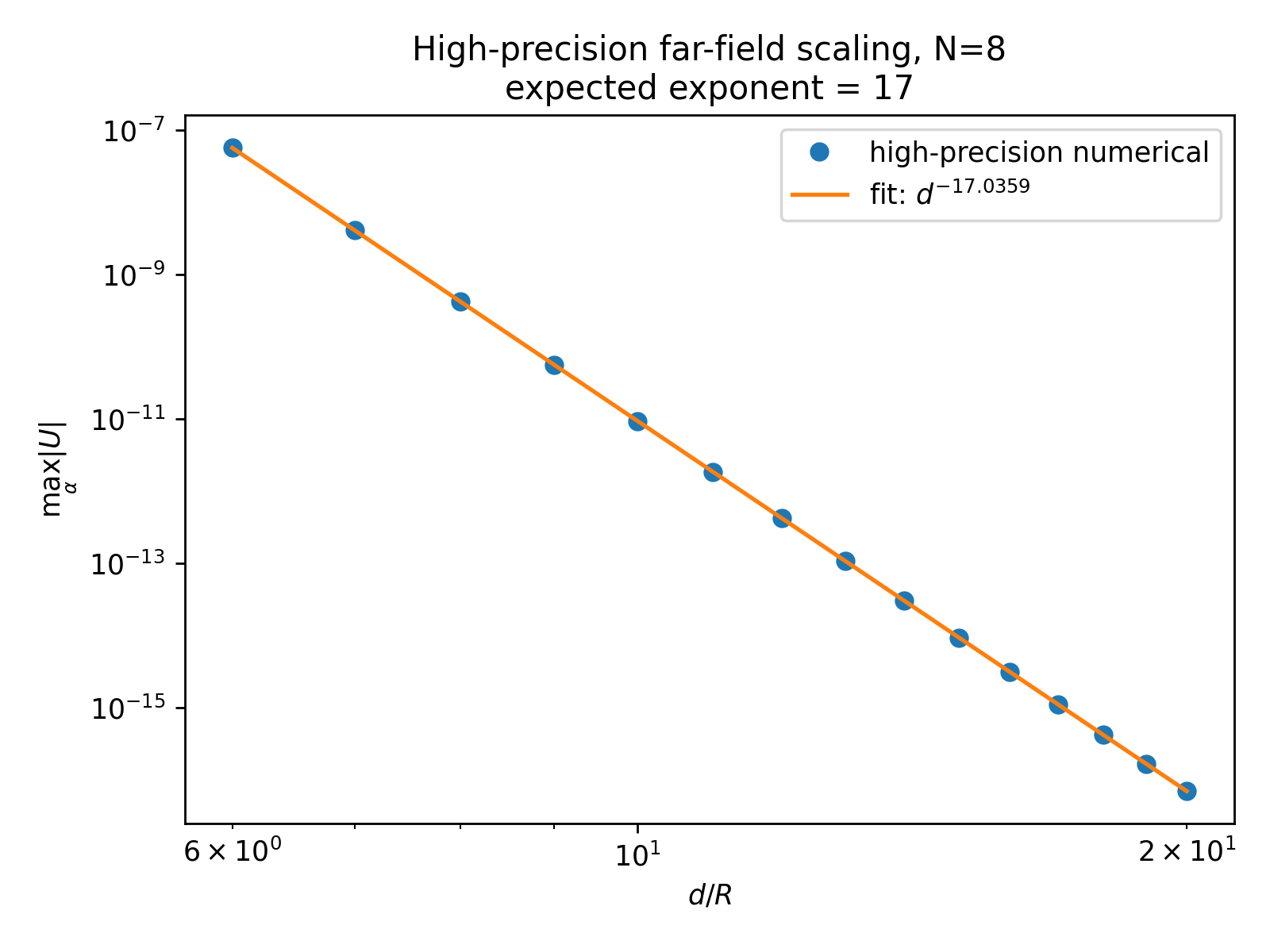}\\[-1mm]
			$N=8$
		\end{minipage}
		
		\vspace{2mm}
		
		\begin{minipage}{0.48\textwidth}
			\centering
			\includegraphics[width=\linewidth]{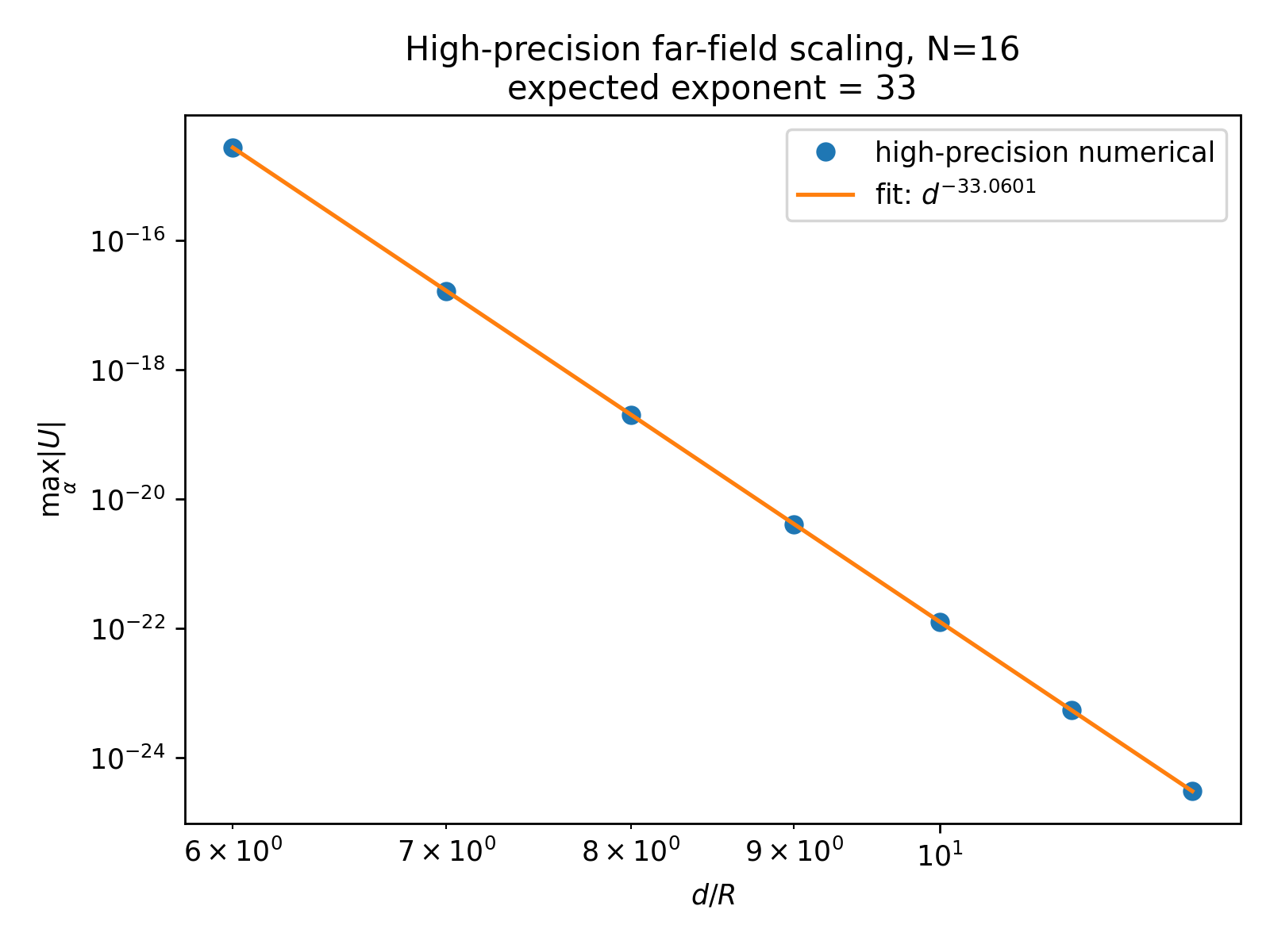}\\[-1mm]
			$N=16$
		\end{minipage}\hfill
		\begin{minipage}{0.48\textwidth}
			\centering
			\includegraphics[width=\linewidth]{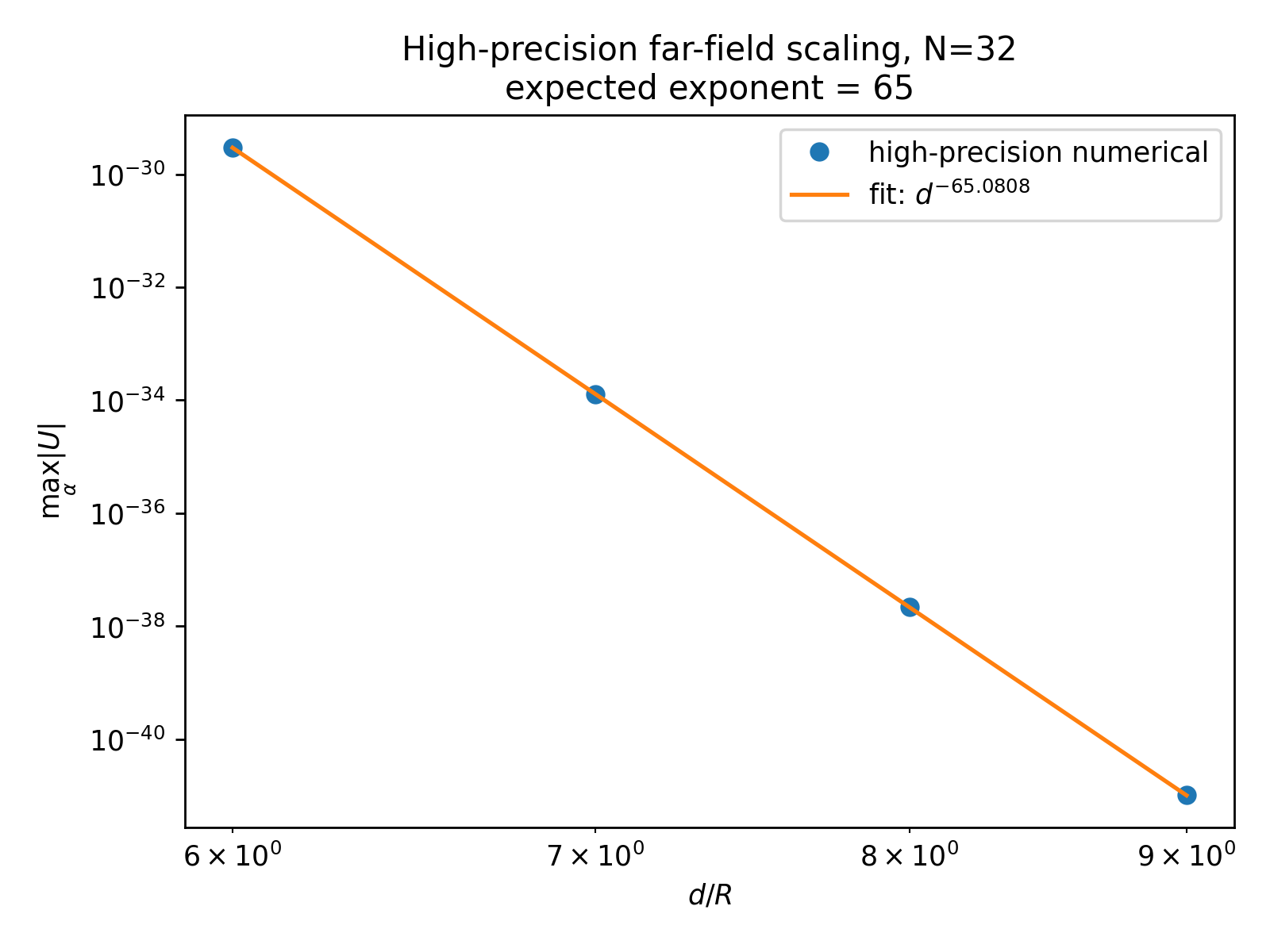}\\[-1mm]
			$N=32$
		\end{minipage}
		\caption{High-precision far-field scaling of $\max_{\alpha}|U_N|$ for $N=4,8,16,$ and $32$. The numerical exponents are in excellent agreement with the analytic prediction $U_N\propto d^{-(2N+1)}$.}
		\label{fig:SI_farfield}
	\end{figure}
	
	The convergence and multipole analyses establish that the pronounced $4+4$ interaction landscape is a property of a discrete spin-toroidal texture. The toroidal moment remains a useful collective descriptor of the internal vortex-like arrangement, but the external magnetostatic interaction is governed by the full higher-multipole content of that arrangement.
}

\section{Molecular Model}

{\color{black}
	\subsection{Computational details for the molecular DFT calculations}
	\label{sec:SI_DFT_methods}
	
	The molecular dimer calculations were performed using noncollinear spin-polarized density functional theory including spin--orbit coupling in VASP. For reproducibility, the complete computational setup used in the revised manuscript is summarized here.
	
	\begin{itemize}
		\item Exchange--correlation functional: PBE
		\item Treatment of the Tb $4f$ electrons: The localized Tb $4f$ states were treated within the DFT+$U$ approach.
		\item Hubbard correction: A Hubbard parameter of $U=7$ eV was applied to the Tb $4f$ states. 
		\item PAW/POTCAR datasets: Projector-augmented-wave (PAW) potentials supplied with VASP were used (C, H, N, O, S, and Tb\_3). 
		\item Plane-wave cutoff: A plane-wave kinetic-energy cutoff of $500$ eV was used. 
		\item Electronic convergence criterion: The electronic self-consistency cycle was converged to
		$10^{-6}$ eV, with a maximum of 400 electronic iterations.
		\item Ionic relaxation/convergence:All molecular orientations were evaluated using single-point
		calculations with fixed atomic geometries taken from the experimental reference stated in main text. 
		\item Dispersion treatment: D3. 
		\item Charge state of the molecular dimer: Neutral. 
		\item Supercell and vacuum conditions: The molecular dimer was placed in an orthorhombic supercell of
		dimensions
		$50.4789\times30.4789\times30.4789~\text{\AA}^{3}$.
		The two molecular units were separated by a center-to-center
		distance of $18.0$ \AA.
		\item $k$-point sampling:  $\Gamma$-point only. 
		\item Noncollinear/SOC settings: Noncollinear spin-polarized calculations including spin--orbit
		coupling were performed with symmetry disabled
		(\texttt{ISYM}=0).
		\item Initial local Tb magnetic moments: The initial local magnetic moment on each Tb site was set to a
		magnitude of $3\,\mu_{\mathrm B}$. The local Tb magnetic moments were constrained to maintain the prescribed vortex-like magnetic configuration during the orientation-dependent total-energy calculations using the constrained-local-moment approach in VASP. 
	\end{itemize}
	
	These DFT energies are total molecular energies and are not decomposed into a unique toroidal magnetic contribution. The molecular calculation is therefore used as a qualitative proof-of-concept comparison for orientation-dependent anisotropy, rather than as a quantitative validation of the classical dipole--dipole interaction energy.
}

\begin{figure}[H]
	\centering
	\includegraphics[width=0.85\textwidth]{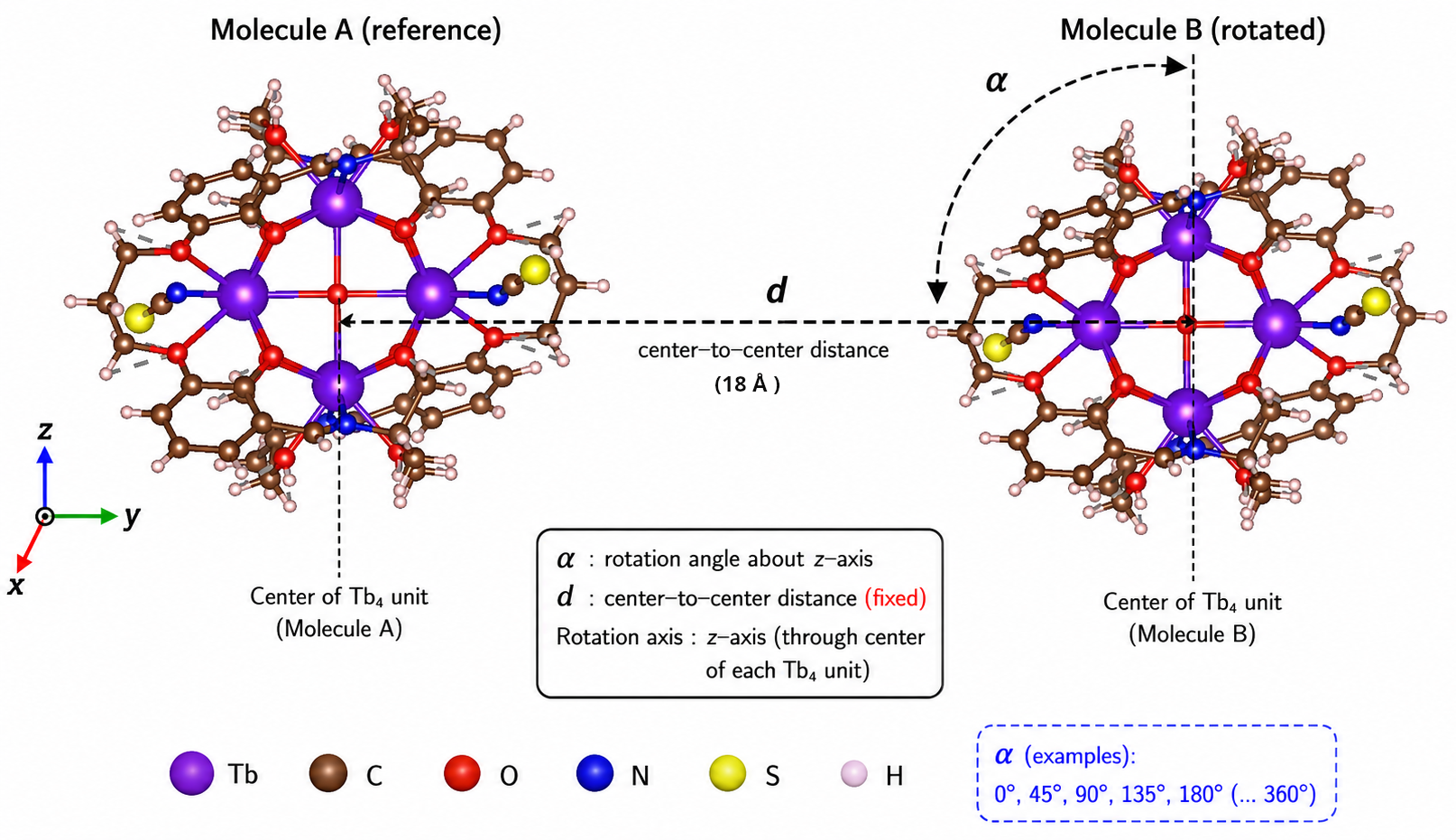}
	\caption{
		Geometry of the Tb$_4$ molecular toroidal dimer used in the orientation-dependent interaction calculations. Two extracted Tb$_4$ molecular fragments are placed at a fixed 18 {\AA} center-to-center separation distance $d$, while one molecular unit is rotated by an angle $\alpha$ with respect to the other. The local Tb moments are initialized in a vortex-like tangential arrangement to mimic the toroidal magnetic configuration. 
	}
	\label{fig:tb4}
\end{figure}

\bibliography{spin-toroidal-moment}
\bibliographystyle{unsrt}

\end{document}